\documentclass[aps,prd,onecolumn,eqsecnum,amsmath,nofootinbib,preprintnumbers]{revtex4}%

\usepackage{color,graphicx,float,subfigure,xcolor}
\usepackage{amsfonts,amssymb,theorem,mathrsfs,times}
\usepackage{bm}
\usepackage{booktabs}
\usepackage{mathtools}
\usepackage{amsfonts,amssymb,theorem,mathrsfs}
\usepackage{dsfont}
\usepackage{setspace}
\usepackage{amsmath}  
\usepackage[colorlinks,linkcolor=blue,anchorcolor=blue,citecolor=blue]{hyperref}   
\usepackage{ulem}   
\usepackage[english]{babel}
\newtheorem{theorem}{Theorem}

{\theorembodyfont{\upshape}
	}
{\theorembodyfont{\upshape}
	}
{\theorembodyfont{\upshape}
	}
{\theorembodyfont{\upshape}
	}
{\theorembodyfont{\upshape}
	}
{\theorembodyfont{\upshape}
	}
\addtocounter{MaxMatrixCols}{10}
\newcommand{\dalm}{\kern1pt\vbox{\hrule height 0.9pt\hbox{\vrule width
			0.9pt\hskip 2.5pt\vbox{\vskip 5.5pt}\hskip 3pt\vrule width
			0.3pt}\hrule height 0.3pt}\kern1pt}

\begin{document}
\thispagestyle{empty}
\title{The pseudospectrum and transient of Kaluza-Klein black holes in Einstein-Gauss-Bonnet gravity}
	
%

\author{ Jia-Ning Chen$^{1,2,3}$\footnote{e-mail address: chenjianing22@mails.ucas.ac.cn}}

\author{ Liang-Bi Wu$^{1,2}$\footnote{e-mail
			address: liangbi@mail.ustc.edu.cn}}

\author{ Zong-Kuan Guo$^{3,2,1}$\footnote{e-mail
			address: guozk@itp.ac.cn}} 
   
\affiliation{${}^1$School of Fundamental Physics and Mathematical Sciences, Hangzhou Institute for Advanced Study, University of Chinese Academy of Sciences, Hangzhou 310024, China}

\affiliation{${}^2$School of Physical Sciences, University of Chinese Academy of Sciences, No.19A Yuquan Road, Beijing 100049, China}

\affiliation{${}^3$CAS Key Laboratory of Theoretical Physics, Institute of Theoretical Physics, 
Chinese Academy of Sciences, Beijing 100190, China}

\date{\today}

\begin{abstract}
    The spectrum and dynamical instability, as well as the transient effect of the tensor perturbation for the so-called Maeda-Dadhich black hole, a type of Kaluza-Klein black hole, in Einstein-Gauss-Bonnet gravity have been investigated in framework of pseudospectrum. We cast the problem of solving quasinormal modes (QNMs) in AdS-like spacetime as the linear evolution problem of the non-normal operator in null slicing by using ingoing Eddington-Finkelstein coordinates. In terms of spectrum instability, based on the generalised eigenvalue problem, the QNM spectrum and $\epsilon$-pseudospectrum has been studied, while the open structure of $\epsilon$-pseudospectrum caused by the non-normality of operator indicates the spectrum instability. In terms of dynamical instability, we introduce the concept of the distance to dynamical instability, which plays a crucial role in bridging the spectrum instability and the dynamical instability. We calculate such distance, named the complex stability radius, as parameters vary. Finally, we show the behaviour of the energy norm of the evolution operator, which can be roughly reflected by the three kinds of abscissas in context of pseudospectrum, and find the transient growth of the energy norm of the evolution operator.
\end{abstract}

\maketitle

\section{Introduction}
The black hole spectroscopy program has made significant progress by utilizing gravitational waves (GWs) generated from the merger of binary systems. These GW signals provide valuable information about the characteristics of the merging systems~\cite{LIGOScientific:2016aoc,LIGOScientific:2016sjg,LIGOScientific:2017bnn,LIGOScientific:2017vwq,KAGRA:2021vkt}. This achievement has sparked the development of various space-based initiatives, including LISA~\cite{LISA:2017pwj}, TianQin~\cite{TianQin:2015yph,Gong:2021gvw}, and Taiji~\cite{Hu:2017mde}. During the ringdown phase, the remnant black hole emits a distinct signal as it dissipates energy and settles into its equilibrium state. This signal is characterized by a discrete collection of damped harmonic oscillations known as quasinormal modes~(QNMs)~\cite{Kokkotas:1999bd,Berti:2009kk,Konoplya:2011qq,Franchini:2023eda}.

With improvement of the accuracy of gravitational waves detection, research has focused on the QNM spectra of black holes in the presence of surrounding matter, since black holes are not isolated but are always accompanied by an astrophysical environment~\cite{Barausse:2014tra,Cannizzaro:2024yee}. The characteristics of QNMs are closely tied to the concept of spectrum instability. Initial studies by Nollert and Price~\cite{Nollert:1996rf,Nollert:1998ys}  have demonstrated that even small perturbations can have a significant impact on the QNM spectrum, leading to what is known as spectrum instability. This discovery highlights the unexpected sensitivity of the QNM spectrum to ``ultraviole'' or small-scale perturbations, challenging the assumption that a good approximation of the effective potential would yield minimal deviations in the observed QNMs.

After extensive surveys, we have identified two main categories of methods used to study spectrum instability. However, it is important to note that there can be some overlap or connections between these two approaches. The first approach involves modifying the effective potential based on various reasons, which can be either physically motivated or artificially constructed. For instance, studies~\cite{Daghigh:2020jyk,Qian:2020cnz} have argued that even minor modifications, such as using a piecewise approximate potential, can lead to non-perturbative changes in the asymptotic behavior of the QNM spectrum. Authors in~\cite{Hirano:2024fgp} utilized the parameterized black hole quasinormal ringdown formalism and observed larger deviations from general relativity in the quasinormal frequencies for higher overtones.


The addition of small wave packets to the existing effective potential is a commonly encountered method in the study of spectrum instability~\cite{Cardoso:2024mrw,Courty:2023rxk,Berti:2022xfj,Cheung:2021bol,Rosato:2024arw,Oshita:2024fzf}. Notably, studies~\cite{Rosato:2024arw,Oshita:2024fzf} focus on the (in)stability of black hole greybody factors and find that these factors are stable observables for detecting gravitational waves, meanwhile Regge Poles are used to construct stable observables in~\cite{Torres:2023nqg} considering the unstable nature of the QNM spectrum. In~\cite{Cardoso:2024mrw}, it is argued that the addition of a small bump $\epsilon V_{\text{bump}}$ cannot be considered as a small perturbation in any sense. According to their perspective, the introduction of any arbitrarily small bump leads to the emergence of a new length scale, which explains the instability of the fundamental mode. Additionally, the previous work approaches the instability of spectra from the perspective of equal perturbation energy~\cite{Cao:2024oud}. More recently, modified PT effective potentials have been considered to study spectrum instability~\cite{Li:2024npg}.


Pseudospectrum analysis~\cite{trefethen2020spectra}, as a second approach, is utilized to investigate the instability of the QNMs spectrum~\cite{Jaramillo:2020tuu,Destounis:2023ruj,Jaramillo:2021tmt}, which originally comes from the field of hydrodynamics~\cite{doi:10.1126/science.261.5121.578}. This method involves analyzing the properties of non-self-adjoint operators in dissipative systems and provides a visual understanding of the instability in the spectrum of such operators~\cite{trefethen2020spectra}. In the context of black holes, pseudospectra have been employed to identify qualitative characteristics that serve as indicators of spectral instability across a range of spacetimes. These investigations not only encompass the findings of ~\cite{Jaramillo:2020tuu}, but also extend them to other types of black holes, including asymptotically flat black holes~\cite{Destounis:2021lum,Cao:2024oud}, asymptotically AdS black holes~\cite{Arean:2023ejh,Cownden:2023dam,Boyanov:2023qqf}, asymptotically dS black holes~\cite{Sarkar:2023rhp,Destounis:2023nmb}, and horizonless compact objects~\cite{Boyanov:2022ark}.

Ref.\cite{Jaramillo:2022kuv} has recently expanded the applications of pseudospectrum in the field of gravitational wave physics. This work focuses on dynamic phenomena, specifically transient growths in linear dynamics and the occurrence of pseudoresonances, which arise from non-self-adjoint operators. While the decaying QNMs of the evolution operator govern the later stages of the linear system's dynamics, its non-self-adjoint nature can lead to unexpected transient growth in the early stages. These transient and pseudoresonant effects are not fully captured by eigenvalues alone. Instead, they are encoded within the comprehensive structure of the pseudospectrum and evaluated using dedicated quantities derived from it. Using the hyperboloidal approach, Ref.\cite{Jaramillo:2022kuv} reveals that there is no early-time transient growth observed in asymptotically flat spacetimes. This result remains unchanged regardless of the specific potential governing the wave dynamics and can be mathematically attributed to the escape of the field through the spacetime boundaries, including the horizon and infinity. The study~\cite{Boyanov:2022ark} investigates the pseudospectrum of the linear operator associated with exotic compact objects (ECOs) featuring reflective surfaces, aiming to gain insights into transient growth. In a more recent work, Ref.\cite{Carballo:2024kbk} explores transient dynamics within the framework of general relativity, including asymptotically dS black holes, asymptotically AdS black holes, and asymptotically flat black holes. The study demonstrates that non-normality gives rise to arbitrarily long-lived sums of short-lived QNMs.


It is worthwhile to investigate pseudospectrum and transient from the perturbation equation within the context of modified gravity theories. The study object is the Maeda-Dadhich black hole~\cite{Maeda:2006iw}, which is a type of Kaluza-Klein (KK) black hole in Einstein-Gauss-Bonnet (EGB) gravity. Importantly, a generalized master equation for tensor perturbations of EGB gravity has been derived in the previous works~\cite{Cao:2021sty,Cao:2023zhy}. We select such a black hole for three primary reasons.  Firstly, from a theoretical perspective, this black hole constitutes an exact vacuum solution within EGB gravity with a cosmological constant that holds a distinctive relationship to the Gauss-Bonnet coupling constant, and exhibits a fascinating trait where the Gauss-Bonnet term behaves akin to a Maxwell source at large distances, while simultaneously regularizing the metric and mitigating the central singularity~\cite{Maeda:2006iw}. Secondly, this choice allows for an excellent model to explore extra dimensions in framework of modified gravity theory. Modified gravity theories containing extra dimensions offer valuable insights into the constraints of Einstein's gravity theory, tackle challenges faced by current models, and play a crucial role in advancing our comprehension of the Universe. EGB gravity whose Lagrangian contains only the linear and quadratic terms of spacetime curvature is widely studied among theoretical physicists. Black holes within high-dimensional spacetimes have attracted considerable attention due to their natural occurrence within string theory and their presence in scenarios involving extra-dimensional brane worlds. The advancement in gravitational wave observations enables the use of QNMs to extract valuable information about the black hole, potentially validating modified gravity theories featuring extra dimensions in future research. Lastly, our decision is influenced by considerations in null slicing calculations. The asymptotic behavior of this specific black hole mirrors a field in Anti-de Sitter (AdS) spacetime. Calculating QNMs by utilizing null slicing in AdS spacetime serves as a significant motivation for our selection. We anticipate that the transient effects observed are not exclusive to general relativity or hyperboloidal slicing, but may also manifest in modified gravity theories and when utilizing null slicing as a framework. Taking into account these factors, we have opted to focus on the Maeda-Dadhich black hole for our study.

We all know that the strong external environment may affect the linear perturbation, making it unstable in dynamics, that is, resulting in an exponentially growing mode. With the energy norm, we define the (complex) stability radius to quantitatively characterize the distance to dynamical instability, and find different parameters correspond to different stability radii. More importantly, we study the transient effect of linear tensor perturbations of Maeda-Dadhich black holes in EGB gravity in terms of the norm of the evolution operator. It is found that the numerical abscissa is larger than zero, which indicated the transient growth will happen. In addition, we also confirmed such growth  by calculating the evolution of the energy norm of a field. This transient growth can also lead to an increase in the maximum value of the waveform, which we illustrate through numerical simulations. 

The remains of this paper are organized as follows. In Sec.\ref{Maeda-Dadhich_black_hole}, we give a very short review of Maeda-Dadhich black hole, meanwhile the Schr\"{o}dinger-like equation is derived. Sec.\ref{set_up} offers some necessary preparations for numerical calculation of pseudospectrum, including ingoing Eddington–Finkelstein coordinates and the associated physical energy norm. In Sec.\ref{spectrum_and_pseudospectrum}, we show the QNM frequencies and the pseudospectrum to convey the properties of spectrum instability. In Sec.\ref{pseudospectrum_and_the transients}, we clarify the relation among pseudospectrum, stability radius and transient effect. Sec.\ref{conclusions_and_discussion} provides the conclusions and discussions. In Appendix.\ref{pseudospectrum_of_the_generalized_eigenvalue_problems}, the definition of the pseudospectrum of a generalized eigenvalue problem is given. In Appendix.\ref{derivation of partail E}, we derive the expression of $\partial_{v}E[\varphi]$. In Appendix.\ref{numerical_approach_of_pseudospectrum}, the numerical approach of pseudospectrum is come up. In Appendix.\ref{convergence_test}, the convergence test of QNMs is shown. In order to obtain the waveform, the numerical method used in this work is appeared at Appendix.\ref{numerical_method}.

\section{The Maeda-Dadhich black hole and the Schr\"{o}dinger-like equation}\label{Maeda-Dadhich_black_hole}

\begin{figure}[htbp]
    \centering
    \includegraphics[width=\textwidth]{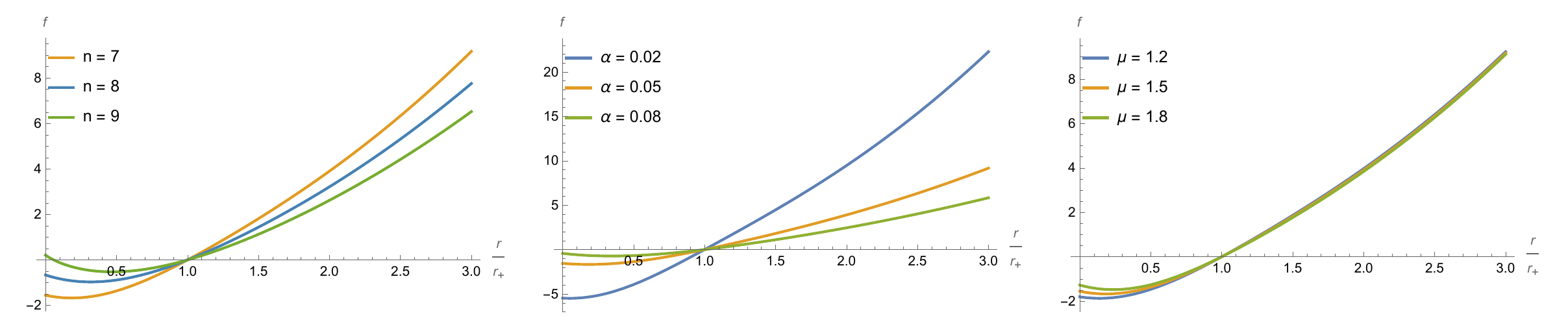}
    \caption{The metric function $f(r)$ varying with the three parameters $n$, $\alpha$ and $\mu$ are depicted by maintaining the event horizon being $r_{+}=1$. The varying parameters corresponding to each color line are indicated in the legends. The benchmark parameters are $n = 7$, $\alpha = 0.05$, and $\mu = 0.5$, which is the orange line among above three panels.}
    \label{fig:Plotfr}
\end{figure}

We begin by introducing a Kaluza-Klein black hole  proposed by Maeda and Dadhich~\cite{Maeda:2006iw}, which is a vacuum solution in Einstein-Gauss-Bonnet gravity theory. The action for $n\ge 5$ in the $n$-dimensional spacetime with a metric $g_{MN}$ is given by 
\begin{eqnarray}
	S=\int \mathrm{d}^nx\sqrt{-g}\Big[\frac{1}{2\kappa_n^2}\Big(R-2\Lambda+\alpha L_{\text{GB}}\Big)\Big]\, ,
\end{eqnarray}
where $\kappa_n$ is the coupling constant of gravity which depends on the dimension of spacetime, $R$ and $\Lambda$ are the $n$-dimensional Ricci scalar and the cosmological constant, respectively. The Gauss-Bonnet term is given by 
\begin{eqnarray}
	L_{\text{GB}}=R^2-4R_{MN}R^{MN}+R_{MNPQ}R^{MNPQ}\, ,
\end{eqnarray}
where the capital letters $\{M,N,P,Q,\cdots\}$ are the indices for the $n$-dimensional spacetime. The symbol $\alpha$ is the coupling constant of the Gauss-Bonnet term, which is identified with the inverse string tension, thereby it's positive definite. The equation of motion is
\begin{eqnarray}\label{EOM}
	G_{MN}+\alpha H_{MN}+\Lambda g_{MN}=0\, ,
\end{eqnarray}
where $G_{MN}$ is the Einstein tensor and $H_{MN}$ is the Gauss-Bonnet tensor. The so-called Maeda-Dadhich black hole spacetime is locally homeomorphic to $M^4\times\mathcal{K}^{n-4}$ with the metric reading as
\begin{eqnarray}
\label{Maeda_Dadhich_black_hole_metric}
    g_{MN}\mathrm{d}x^M\mathrm{d}x^N=-f(r)\mathrm{d}t^2+\frac{1}{f(r)}\mathrm{d}r^2+r^2(\mathrm{d}\theta^2+\sin^2\theta\mathrm{d}\phi^2)+r_0^2\gamma_{ij}\mathrm{d}x^i\mathrm{d}x^j\, ,
\end{eqnarray}
where $r_0$ satisfies  
\begin{eqnarray}
    r_0^2=-2K\alpha(n-4)(n-5)\, ,   
\end{eqnarray}
and $\gamma_{ij}$ is unit metric on the constant curvature manifold $\mathcal{K}^{n-4}$ with sectional curvature $K=-1$. The metric function $f(r)$ in Eq.(\ref{Maeda_Dadhich_black_hole_metric}) is written as
\begin{eqnarray}\label{metric_function_r}
    f(r)=1+\frac{r^2}{2(n-4) \alpha}\Bigg\{1 -\Big[1-\frac{2 n-11}{3(n-5)}+\frac{4(n-4)^2 \alpha^{3 / 2} \mu}{r^3}-\frac{4(n-4)^2 \alpha^2 q}{r^4}\Big]^{1 / 2}\Bigg\}\, ,
\end{eqnarray}
in which $\mu$ and $q$ are arbitrary dimensionless constants with $\mu$ referring to the mass of the central object and $q$ being the chargelike parameter. For simplicity, $\mu\ge0$ and $q\le0$ will be assumed in the followings to ensure that the metric function $f(r)$ exists at least one zero point. One can find some other reasons for choosing such ranges in~\cite{Cao:2023zhy}. Without loss of generality, the units can be fixed, where the event horizon radius is at $r_{+} = 1$. It means that we have $f(1)=0$. For convenience, the constant $q$ will be represented by other parameters via
\begin{eqnarray}
    q&=&\frac{1}{12 \alpha ^2 (n-5) (n-4)^2}\Big[-2 n+11+\alpha(-12 n^2+108 n-240)\nonumber\\
    &&+\alpha ^{3/2}(-960 \mu +12 \mu  n^3-156 \mu  n^2+672 \mu  n)+\alpha ^2(-12 n^3+156 n^2-672 n+960)\Big]\, .
\end{eqnarray}
In other words, the four parameters $n$, $\alpha$, $\mu$ and $q$ are not independent due to the requirement $f(1)=0$, and the metric function $f(r)$ changing with different parameters are shown in Fig.\ref{fig:Plotfr}. From now on, the parameter $\alpha$ should be referred to be dimensionless. 

The master equation of the tensor perturbation based on the metric Eq.(\ref{Maeda_Dadhich_black_hole_metric}) have been given in~\cite{Cao:2021sty}, after the separation of the variables~\cite{Cao:2023zhy,Dotti:2004sh}, the Klein-Gordan-like equation of a scalar field $\Phi$ on the manifold $M^{4}$ can be obtained
\begin{eqnarray}\label{KK_equation}
    \left[\frac{4 n-22}{(n-4)(n-5)} g^{ab}-4 \alpha \cdot{ }^4 G^{a b}\right] D_a D_b \Phi+\left[\frac{2+\gamma}{(n-4)(n-5)}{}^4 R+\frac{3(n-6)(2+\gamma)}{\alpha(n-4)^2(n-5)^2}\right] \Phi=0 \, , 
\end{eqnarray}
where ${}^{4}R$, ${}^4G^{ab}$ are the Ricci scalar and the Einstein tensor on the manifold $M^{4}$, respectively. In fact, the scalar $\Phi$ is the amplitude of the characteristic field $\bar{h}_{ij}$ corresponding to the characteristic value $\gamma$. One can refer to~\cite{Cao:2021sty,Cao:2023zhy} for more detailed processes of the above equation. It should be pointed out here that the constant $\gamma$ appeared in Eq.(\ref{KK_equation}) is the eigenvalue of the Laplace-Beltrami operator acting on the symmetric, transverse, and traceless (STT) tensor fields with rank $2$ on the $(n-4)$-dimensional hyperbolic space $\mathcal{K}^{n-4}$. According to the reference~\cite{Camporesi:1994ga} [see Eq.(2.4) therein], we can obtain 
 \begin{eqnarray}
     \gamma=-\Big[\zeta^2+\Big(\frac{n-5}{2}\Big)^2+2\Big]\, ,
 \end{eqnarray}
 with $\zeta^2$ is continuous and positive. Thus from this perspective, the constant $\gamma$ is limited by $\gamma<-(n-5)^2/4-2$. 
 
Separating the variable $\Phi(t,r,\theta,\phi)$ as
\begin{eqnarray}\label{separating_variables}
	\Phi(t,r,\theta,\phi)=S(r)\varphi(t,r)Y_{lm}(\theta,\phi)\, ,
\end{eqnarray}
where $Y_{lm}(\theta,\phi)$ is the spherical harmonics, we obtain a wave equation with a potential $V(r)$ as follows
\begin{eqnarray}\label{Schrodinger_equation}
    \Big[-\frac{\partial^2}{\partial t^2}+\frac{\partial^2}{\partial r_{\star}^2}-V(r)\Big]\varphi(t,r)=0\, .
\end{eqnarray}
For Eq.(\ref{separating_variables}) and Eq.(\ref{Schrodinger_equation}), the scaling function $S(r)$ and the effective potential $V(r)$ can be found in~\cite{Cao:2023zhy}, $V(r)$ is a function of $n$, $l$, $\alpha$, $\mu$, and $\gamma$, and we omit the symbols of these parameters in the expression $V(r)$. The tortoise coordinate $r_{\star}$ is defined as $\mathrm{d}r_{\star}=\mathrm{d}r/f(r)$ here. When $r\to+\infty$, we obtain the asymptotic behavior of the effective potential as follows
\begin{eqnarray}
    \label{eq: effective_potential_spatial_infinity}
    V(r)\sim V_0(\alpha,n,\gamma)r^2\, ,
\end{eqnarray}
where we only keep the leading order term. The stability requirement demands that $V(r)$ tends towards positive infinity, i.e., $V_0(\alpha,n,\gamma)>0$~\cite{1995AmJPh63256B}, which also limits the range of $\gamma$~\cite{Cao:2023zhy}. Combining two limitations of $\gamma$, the ranges of $\gamma$ in different spacetime dimensions are shown in Tab.\ref{tab:Limitations of gamma}.
\begin{table}[htbp]
    \centering
    \begin{tabular}{l c}
    \toprule[1pt]
    Spacetime dimension $n$ &  Value range of $\gamma$ \\
     \hline 
     $n=6$ & $\backslash$ \\
     $n = 7$ & $ -4.82843 < \gamma < -3 $   \\
     $n \ge 8$ & $\gamma < -(n-5)^2/4-2 $ \\
    \bottomrule[1pt]
    \end{tabular}
    \caption{The value ranges of $\gamma$ with respect to different spacetime dimensions $n$, which are solved by two limitations $\gamma<-(n-5)^2/4-2$ and $V_0(\alpha,n,\gamma)>0$. The symbol $\backslash$ represents the empty set.}
    \label{tab:Limitations of gamma}
\end{table}

\begin{figure}[htbp]
    \centering
    \includegraphics[width = \textwidth]{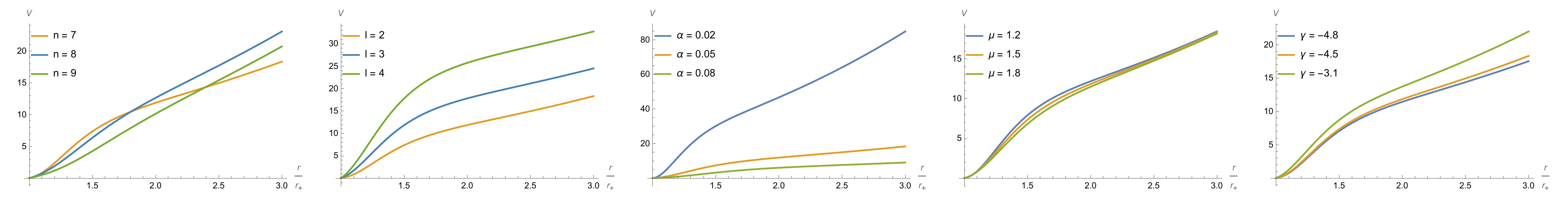}
    \caption{The effective potential $V(r)$ varying with different parameters $n$, $l$, $\alpha$, $\mu$ and $\gamma$ are depicted by maintaining the event horizon be $r_{+}=1$. The varying parameters corresponding to each color line are indicated in the legends. The benchmark parameters are $n=7$, $l=2$, $\alpha=0.05$, $\mu=1.5$ and $\gamma =-4.5$, which is the orange line among above five panels.}
    \label{fig:PlotVr}
\end{figure}

Furthermore, Fig.\ref{fig:PlotVr} shows the variation of the effective potential $V(r)$ with different parameters $n$, $l$, $\alpha$, $\mu$, and $\gamma$. We choose these parameters for four reasons: (i) the ranges of $q$ and $\mu$ are selected as $q \le 0$ and $\mu \ge 0$, (ii) the potential $V(r)$ should remain positive from the event horizon to infinity, (iii) the effective AdS radius~\cite{Cao:2023zhy} calculated with these parameters can be compared with the event horizon $r_{+}$, and it should also ensure that $r_{+}$ is the outermost event horizon and $r_{+} = 1$ if there exists another horizon $r_{-}$, (iv) the constant $\gamma$ should be chosen as illustrated in Tab.\ref{tab:Limitations of gamma}. The ranges for the five parameters are fixed as $n =[7,9]$, $l$ starting from $l=2$, $\alpha = [0.02,0.08]$, $\mu = [1.2,1.8]$, and $\gamma = [-4.8,-3.1]$ in the overall subsequent calculations to investigate physical phenomena, although they are not unique ranges for these parameters. There are two sets of parameters that need to be clarified. One is the set of benchmark parameters, which are fixed as $n=7$, $l = 2$, $\alpha = 0.05$, $\mu = 1.5$, and $\gamma = -4.5$. In these parameters, the effective AdS radius is $\ell_{\text{AdS}}= 1.01206$, which is comparable to the event horizon. The other set is when the spacetime dimension is $n=9$, and currently $\gamma$ is limited to $\gamma < -6$ due to condition (iv). Therefore, $\gamma = -6.5$ is fixed when $n=9$.

\section{The set up}\label{set_up}
\subsection{The establishment of eigenvalue problems}
In this subsection, we utilize the ingoing Eddington-Finkelstein (EF) coordinates to cast the problem of solving for the QNM frequencies as an eigenvalue problem~\cite{Cownden:2023dam}. There are two reasons for choosing this coordinate system. First, these coordinates exhibit regularity at the black hole horizon and the AdS boundary is allowed to arrive at. By utilizing these coordinates, we can impose the necessary boundary conditions at the extremities of the physical domain, which correspond to the two ends of $v=\text{constant}$ surfaces. However, it is important to note that this approach is not feasible in asymptotically flat spacetimes, where surfaces of constant $v$ lead to past null infinity $\mathcal{J}^{-}$ instead of future null infinity $\mathcal{J}^{+}$. The motivation for proposing hyperboloidal coordinates is to address this issue~\cite{Zenginoglu:2007jw,Zenginoglu:2011jz,Zenginoglu:2024bzs,Jaramillo:2020tuu,PanossoMacedo:2023qzp}. However, in our current analysis, it is found that hyperboloidal coordinates are not necessary and the ingoing EF coordinates are simpler to work with. The drawback of using these coordinates is that it leads to a generalized eigenvalue problem, which has a different structure compared to the standard eigenvalue problem. The second reason for choosing the ingoing EF coordinates is based on the discussion regarding the structural aspects of the AdS black hole pseudospectrum, as presented by Boyanov \textit{et al.}~\cite{Boyanov:2023qqf}. Their work suggests that the convergence properties of the pseudospectrum are more favorable in such case. However, the discussion on the convergence of the pseudospectrum goes beyond the scope of our work.

Given the two aforementioned reasons, it is reasonable for us to use the ingoing EF coordinates to handle our problem. For convenience, we can compactify the radial coordinate $r$ by introducing $z=1/r$. By doing so, the black hole event horizon is located at $z=1$ and the AdS boundary is located at $z=0$. Therefore, combining the ingoing EF coordinate and the compactification coordinate $z$, we have the coordinate transformation, which is given by
\begin{eqnarray}\label{ingoing_EF_coordinate}
    v=t+ r_{\star}\, ,\quad \frac{\mathrm{d} r_{\star}}{\mathrm{d} r}=\frac{1}{f(r)}\, , \quad z=\frac{1}{r}\, .
\end{eqnarray}
From the above coordinate transformations, one obtains the metric in ingoing EF coordinates on the  four dimensional manifold $M^4$
\begin{eqnarray}
    \mathrm{d}s^2=-f(z)\mathrm{d}v^2-\frac{2}{z^2}\mathrm{d}v\mathrm{d}z + \frac{1}{z^2}( \mathrm{d}\theta^2 + \sin^{2}\theta \mathrm{d}\phi^2 )\, ,
\end{eqnarray}
where
\begin{eqnarray}
    f(z)=1+\frac{1}{2(n-4) \alpha z^2}\left\{1 -\Big[1-\frac{2 n-11}{3(n-5)} + 4z^3(n-4)^2 \alpha^{3 / 2} \mu -  4z^4(n-4)^2 \alpha^2 q \Big]^{1 / 2}\right\}\, .
\end{eqnarray}

By using the ingoing EF coordinates, Eq.(\ref{Schrodinger_equation}) becomes\footnote{For convenience, we write $V(r(z))$ into $V(z)$.}
\begin{eqnarray}\label{Schrodinger_equation_v_z_varphi}
    \Big[-2z^2f(z)\partial_{vz}+z^2f(z)\partial_z(z^2f(z)\partial_z)-V(z)\Big]\varphi(v,z)=0\, .
\end{eqnarray}
In order to regularize the solution of Eq.(\ref{Schrodinger_equation_v_z_varphi}) at the AdS boundary $z=0$, we proceed to redefine
\begin{eqnarray}\label{redefine}
    \varphi(v,z) = z^{\lambda} \psi(v,z) \, , 
\end{eqnarray}
where the index $\lambda>0$ is confirmed as~\cite{Cao:2023zhy} 
\begin{eqnarray}\label{index_lambda}
    \lambda = \frac{1}{2} \Bigg\{1+\sqrt{ 1 + \frac{ 16(n-4)^{2}\alpha^{2}V_{0} }{ \Big[ 1 - \sqrt{\frac{n-4}{3(n-5)}} \Big]^{2} } }\Bigg\}  \, .
\end{eqnarray}
By substituting Eq.(\ref{redefine}) into Eq.(\ref{Schrodinger_equation_v_z_varphi}), we have 
\begin{eqnarray}\label{Schrodinger_equation_v_z_psi}
    &&z^{\lambda}\Bigg\{-2\lambda zf(z)\partial_v\psi(v,z)-2z^2f(z)\partial_{vz}\psi(v,z)+z^4f^2(z)\partial_z^2\psi(v,z)\nonumber\\
    &&+\Big[2(\lambda+1)z^3f^2(z)+z^4f(z)f^{\prime}(z)\Big]\partial_z\psi(v,z)\nonumber\\
    &&+\Big[\lambda(\lambda+1)z^2f^2(z)+\lambda z^3f(z)f^{\prime}(z)-V(z)\Big]\psi(v,z)\Bigg\}=0\, ,
\end{eqnarray}
where the prime denotes derivative with respect to $z$. Therefore, through multiplying the factor $\Big[z^{\lambda+1}f(z)\Big]^{-1}$ into Eq.(\ref{Schrodinger_equation_v_z_psi}), we obtain the equation of evolution in terms of $\psi(v,z)$, which is given by\footnote{We may call $H=L_2^{-1}L_1$ is the Hamiltonian of our present system.}
\begin{eqnarray}\label{equation_of_evolution}
    L_{2}\partial_{v} \psi(v,z) = L_{1}\psi(v,z)\, ,
\end{eqnarray}
where the operators $L_1$ and $L_2$ are satisfied with
\begin{eqnarray}\label{L_1}
    L_{1} = z^{3}f(z)\partial_z^2 + \Big[2(\lambda+1)z^2f(z)+z^3f^{\prime}(z)\Big]\partial_z + \Big[\lambda(\lambda+1)zf(z)+\lambda z^2f^{\prime}(z)-\frac{V(z)}{zf(z)}\Big]\, ,
\end{eqnarray}
\begin{eqnarray}\label{L_2}
    L_2=2z\partial_z+2\lambda\, .
\end{eqnarray}

In order to obtain QNM frequencies, we perform a Fourier transform like $\psi(v,z)\sim e^{i\omega v}\psi(z)$, then we are going to get a generalised eigenvalue problem
\begin{eqnarray}\label{generalised_eigenvalue_problem}
      (L_{1}-i\omega L_{2})\psi(z)=0\, .
\end{eqnarray}
By using the Chebyshev grid to discretize two operators $L_1$ and $L_2$ , one gets its matrix versions $\mathbf{L}_1$ and $\mathbf{L}_2$. The generalised eigenvalues $\omega$ (QNM frequencies) of such matrix pencil are derived from 
\begin{eqnarray}
    \text{det}(\mathbf{L}_1-i\omega\mathbf{L}_2)=0\, .
\end{eqnarray}
The pseudospectrum is also based on Eq.(\ref{generalised_eigenvalue_problem}). One can find more details in terms of the pseudospectrum of such generalized eigenvalue problems with matrix version in Appendix.\ref{pseudospectrum_of_the_generalized_eigenvalue_problems}. So far, we have completed the establishment of the eigenvalue problem. Finally, to increase readability, we give the reason for choosing such $\lambda$. Note that $f\sim f_0/z^2$ and $V(z)\sim V_0/z^2$ as $z\to0$, in which 
\begin{eqnarray}
    f_0=\frac{1}{2(n-4)\alpha}\Big[1-\sqrt{\frac{n-4}{3(n-5)}}\Big]\, .
\end{eqnarray}
It is easy to find that the first two terms of $L_1$ and the two terms of $L_2$ are all regular in the interval $[0,1]$. The third term in $L_1$ is given by
\begin{eqnarray}\label{third_term}
    \lambda(\lambda+1)zf(z)+\lambda z^2f^{\prime}(z)-\frac{V(z)}{zf(z)}=\frac{1}{z}\Big[\lambda(\lambda-1)f_0-\frac{V_0}{f_0}\Big]+\mathcal{O}(z)\, ,\quad z\to0\, .
\end{eqnarray}
In order to keep the above term regular, the coefficient of $1/z$ has to be vanished. Therefore, the index $\lambda>0$ should be chosen as Eq.(\ref{index_lambda}). 

\subsection{The energy norm}
In order to quantify the size of perturbations, we should define an appropriate norm $\lVert\cdot\rVert_E$, in which the energy norm is more natural. As is well known or empirically speaking, a wide class of perturbations around a black hole can be reduced into such form like Eq.(\ref{Schrodinger_equation}). Such equation has $1+1$ covariant form
\begin{eqnarray} \label{EOM of phi}
    \eta^{\alpha\beta}\nabla_\alpha\nabla_\beta\varphi-V\varphi=0\, ,
\end{eqnarray}
where $\eta_{\alpha\beta}$ is an effective metric with ingoing EF coordinates ($v$, $r_{\star}$) in $2$-dimensional Minkowski spacetime. On the other hand, it is easy to find that Eq.(\ref{EOM of phi}) stems from the action
\begin{eqnarray}\label{effective_action}
    S\sim-\frac{1}{2}\int\mathrm{d}^2y\sqrt{-\eta}\Big(\eta^{\alpha\beta}\nabla_\alpha\bar{\varphi}\nabla_\beta\varphi+V\bar{\varphi}\varphi\Big)\, .
\end{eqnarray}
The energy momentum tensor of the scalar field $\varphi$ from the action (\ref{effective_action}) reads as \cite{Jaramillo:2020tuu,Cownden:2023dam} 
\begin{eqnarray}
    T_{\alpha\beta} = \frac{1}{2} \nabla_{\alpha}\bar{\varphi}\nabla_{\beta}\varphi + \frac{1}{2} \nabla_{\alpha}\varphi \nabla_{\beta}\bar{\varphi} - \frac{\eta_{\alpha\beta}}{2}( \nabla_{\mu}\varphi \nabla^{\mu}\bar{\varphi} + V\varphi\bar{\varphi} ) \, ,
\end{eqnarray} 
where the bar means complex conjugation. There are two Noether conserved quantities associated with this action (\ref{effective_action}). To be specific, energy is the invariance quantity with respect to the time translation, and the other Noether charge $Q$ is related to $U(1)$ gauge transformation. Since we are focusing on the null slicing, then  the energy $E[\varphi]$ on the null slicing can be written as
\begin{eqnarray} \label{energy_at_a_null_hypersurface}
    E[\varphi]= \int_{\Sigma_{v}} T_{\alpha\beta} \xi^{\alpha} \kappa^{\beta} \mathrm{d}\Sigma_{v}=\frac{1}{2} \int_{a}^{b} \mathrm{d}r_{\star}\Big(\partial_{r_{\star}}\varphi \partial_{r_{\star}}\bar{\varphi} + V\varphi\bar{\varphi}\Big) \, ,
\end{eqnarray}
where $\xi=\partial_{v}$ is the Killing vector, $\kappa=-\partial_{r_{\star}}$ is the normal vector to the null hypersurface $\Sigma_{v}$, and $a$ and $b$ are the lower and upper boundary in tortoise coordinate $r_{\star}$. The derivative of $E[\varphi]$ with respect to $v$ is given by
\begin{eqnarray} \label{partial energy v}
     \partial_{v}E[\varphi]= - \frac{1}{4}|\partial_{r_{\star}}\varphi|^{2} \Big|^{b}_{a}+\frac{1}{4}V|\varphi|^{2} \Big|^{b}_{a} + \int_{a}^{b} \mathrm{d}r_{\star} \Big( \frac{1}{2}V\partial_{v}|\varphi|^{2} - \frac{1}{4} \frac{\mathrm{d}V}{\mathrm{d}r_{\star}}|\varphi|^{2} \Big) \, .
\end{eqnarray}
The above equation is deducted from Eqs.(\ref{energy_at_a_null_hypersurface}) by using Eq.(\ref{EOM of phi}) with coordinate $(v, r_{\star})$ and integration by parts, and one can find more details in Appendix.\ref{derivation of partail E}. It should be emphasized that the sign of $\partial_{v}E[\varphi]$ can not be identified directly, which is pretty different from the hyperboloidal case~\cite{Carballo:2024kbk}. Furthermore, the energy norm of $\psi$ based on Eq.(\ref{energy_at_a_null_hypersurface}) is given by~\cite{Cownden:2023dam,Jaramillo:2020tuu}
\begin{eqnarray}\label{energy_norm}
\lVert\psi\rVert_E^2\equiv\langle\psi,\psi\rangle_E&=&\frac12\int_{\Sigma_v}\Big(\partial_{r_{\star}}\bar{\varphi}\partial_{r_{\star}}\varphi+V\bar{\varphi}\varphi\Big)\mathrm{d}r_{\star}\nonumber\\
    &=&\frac{1}{2}\int_0^1\Big[z^{2\lambda+2}f(z)(\partial_z\bar{\psi})(\partial_z\psi)+\lambda z^{2\lambda+1}f(z)\psi\partial_z\bar{\psi}+\lambda z^{2\lambda+1}f(z)\bar{\psi}\partial_z\psi\nonumber\\
    &&+z^{2\lambda-2}\Big(\lambda^2z^2f(z)+\frac{V(z)}{f(z)}\Big)\bar{\psi}\psi\Big]\mathrm{d}z\, .
\end{eqnarray}
In the second line, we have used Eq.(\ref{ingoing_EF_coordinate}) and Eq.(\ref{redefine}). The upper and lower limits of integration come from the range of compactified coordinate $z$. Since one can proof that $V(z=1)=0$~\cite{Cao:2023zhy}, the definition of the inner product
\begin{eqnarray}\label{scalar_product}
    \langle\psi_1,\psi_2\rangle_E&=&\frac{1}{2}\int_0^1\Big[z^{2\lambda+2}f(z)(\partial_z\bar{\psi}_1)(\partial_z\psi_2)+\lambda z^{2\lambda+1}f(z)(\psi_2\partial_z\bar{\psi}_1+\bar{\psi}_1\partial_z\psi_2)\nonumber\\
    &&+z^{2\lambda-2}\Big(\lambda^2z^2f(z)+\frac{V(z)}{f(z)}\Big)\bar{\psi}_1\psi_2\Big]\mathrm{d}z \, .
\end{eqnarray}
is regular for $z\in[0,1]$, which is important for numerical calculation. When $\psi_1=\psi_2=\psi$, Eq.(\ref{scalar_product}) reduces into Eq.(\ref{energy_norm}). One can see~\cite{Gasperin:2021kfv} for further discussions of the inner product. In the Appendix.\ref{numerical_approach_of_pseudospectrum}, we provide a detailed introduction to the numerical
technique of using the energy norm defined above to determine the pseudospectrum, which in turn will inform us of the instability of the QNM frequencies.

\section{The results of the spectrum and the pseudospectrum}\label{spectrum_and_pseudospectrum}

In following sections, we will analyze the properties of linear perturbation dynamics in Maeda-Dadhich black hole in view of pseudospectrum framework. The results of the spectrum and the pseudospectrum are shown in this section. One can find the definition of $\epsilon$-pseudospectrum in Appendix.\ref{pseudospectrum_of_the_generalized_eigenvalue_problems} for our generalised eigenvalue problem. Technically, based on the spectral method, the non-normal differential operator that describes the linear system will becomes a $(N+1) \times (N+1)$ discrete matrix through Chebyshev ploynomials, if we choose that the resolution is $N$. Besides, the invariant subspace method is exploited to improve the computational efficiency for pseudospectrum, which has already been used in~\cite{Cao:2024oud}. 

\begin{figure}[htbp]
    \centering
    \includegraphics[width = 0.7\textwidth]{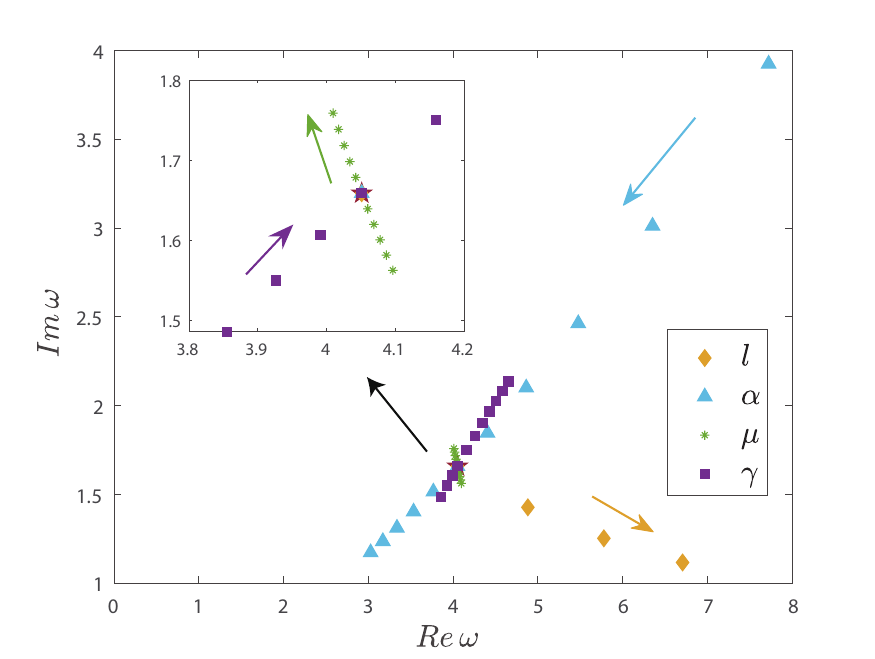}
    \caption{The fundamental mode changes with different parameters $(l, \alpha, \mu, \gamma)$ with numerical resolution $N=100$ and $100$ digits of precision. The benchmark parameters are $n=7$, $l = 2$, $\alpha = 0.05$, $\mu = 1.5$, $\gamma = -4.5$, and the corresponding fundamental mode is represented by $\star$ in the panel and the amplified subgraph. The ranges of parameters used are: $l = [2,5]$, $\alpha = [0.02,0.08]$, $\mu = [1.2, 1.8]$ and $\gamma = [-4.8,-3.1]$, for last three parameters we select $11$ discrete data points. Each color and shape of the points represents each parameter, as illustrated in the legend.}
    \label{fig:fundamentalMOde changes with parameters}
\end{figure}

\begin{figure}[htbp]
    \centering
    \includegraphics[width = 0.45\textwidth]{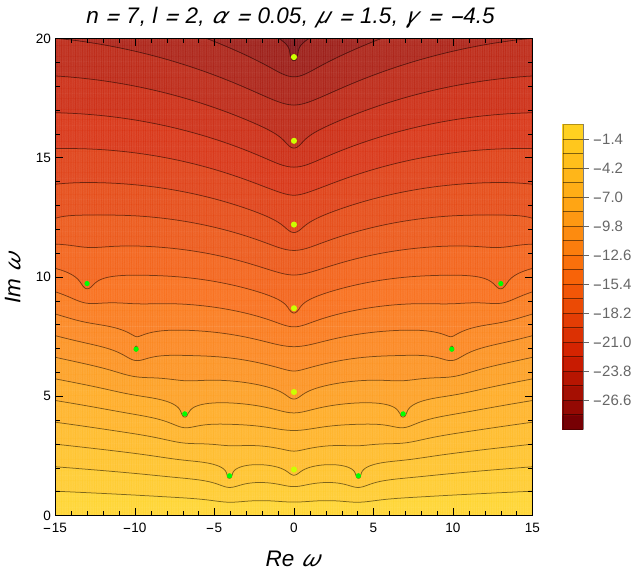}
    \includegraphics[width = 0.45\textwidth]{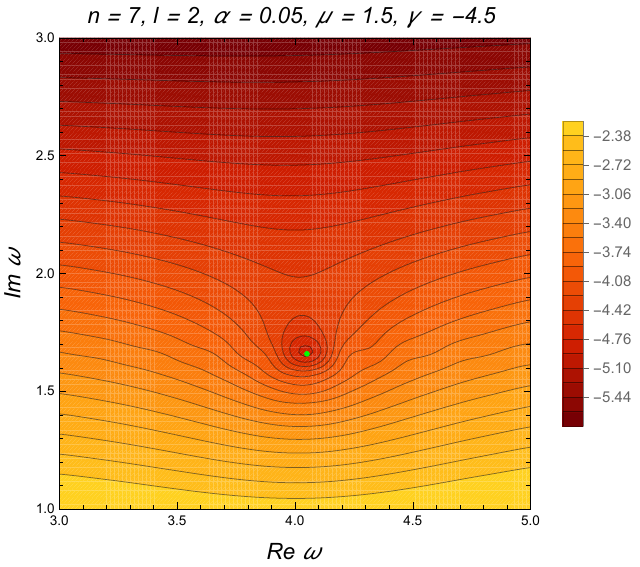}
    \caption{The results of spectrum~(QNM frequencies) and $\epsilon$-pseudospectrum with resolution $N = 100$ and 100 digits of precision. The parameters are the mentioned benchmark parameters, i.e. $n=7$, $l=2$, $\alpha=0.05$, $\mu=1.5$ and $\gamma=-4.5$. The contour lines in both panels denote the boundary of $\epsilon$-pseudospectrum sets given the perturbation size $\epsilon$, whose values are $\log_{10}\epsilon$ and are shown in the legends. The yellow and green points in the left panel are the QNM frequencies in complex plane. The right panel shows $\epsilon$-pseudospectrum around the fundamental mode $\omega_{1} = 4.051108+1.659013i$. Note that the color in the right enlarged panel is inconsistent with the left original panel. One should refer to the actual legend on the right side of each panel.}
    \label{fig:pseudospectrum}
\end{figure}

The fundamental mode varying with different parameters is shown in Fig.\ref{fig:fundamentalMOde changes with parameters}, and the spectrum and pseudospectrum are displayed in Fig.\ref{fig:pseudospectrum} with benmark parameters, where we solve them with resolution $N = 100$ and $100$ digits of precision. The convergence of such eigenvalues in view of the generalised eigenvalue problem are discussed in detail in Appendix.\ref{convergence_test}. 

In Fig.\ref{fig:fundamentalMOde changes with parameters}, we depict the variation of the fundamental mode with different parameters in the Maeda-Dadhich black hole. The ranges of each parameter are specified in Sec.\ref{Maeda-Dadhich_black_hole}. Each color and shape of the points in the panel represent a different parameter, as indicated in the legend. For example, the blue triangle points represent the fundamental mode variation with respect to $\alpha$, and the direction of the colored arrow indicates the trend of the fundamental mode as the corresponding parameter increases. In general, a set of parameters $(n, l, \alpha, \mu, \gamma)$ determines the effective potential, which in turn affects the fundamental mode. The fundamental mode corresponding to the benchmark parameters is located at the intersection of various color and shape points. We amplify the data around this intersection in the subgraph indicated by the black arrow. As $\alpha$ and $\mu$ increase, the effective potential is depressed, resulting in a smaller real part of the fundamental mode. For $\alpha$, the imaginary part of the fundamental mode becomes smaller, while for $\mu$, it becomes larger. On the other hand, as $l$ and $\gamma$ increase, the effective potential is lifted, leading to a larger real part of the fundamental mode. For $l$, the imaginary part of the fundamental mode becomes smaller, while for $\gamma$, it becomes larger. Especially, it is found that for spacetime dimensions $n=8$ and $n=9$, the fundamental modes are both located on the imaginary axis, corresponding to $1.523601i$ and $1.788444i$, respectively, which are not shown in the Fig.\ref{fig:fundamentalMOde changes with parameters}. These results are consistent with the outcomes obtained by using asymptotic iteration method~\cite{Cao:2023zhy}.

In Fig.\ref{fig:pseudospectrum}, we present the spectrum and pseudospectrum with the benchmark parameters. In the left panel, the green and yellow points represent the QNM frequencies in the complex plane. The family of yellow points is located on the imaginary axis, which is meaningful and corresponds to physical modes. However, in flat spacetime, these points are introduced due to the discretization process and are considered meaningless. The right panel shows the $\epsilon$-pseudospectrum around the fundamental mode\footnote{The overtone number is denoted by $\bar{n}$ and it start counting from $1$, the fundamental mode is defined by the smallest value of imaginary part of overall eigenvalues, including eigenvalues on the imaginary axis.}  $\omega_{1}=4.051108+1.659013i$. The open sets formed by the contour lines of the $\epsilon$-pseudospectrum indicate spectrum instability. Specifically, it implies that the QNM frequencies can migrate a further distance in the complex plane than what is possible under a given perturbation size. This property is intrinsic to non-normal operators and is different from the case of normal operators, where the contour lines of the $\epsilon$-pseudospectrum form closed sets. It is worth noting that this situation is different from the case of the hyperboloidal coordinates in global AdS$_{4}$~\cite{Cownden:2023dam}, where the pseudospectrum structure consists of perfectly closed spherical contour lines, indicating the stability of the spectrum.

It is worth mentioning that there are two kinds of instability in view of pseudospectrum framework, one is the spectrum instability based on the spectrum method, which is reflected by the open structure of $\epsilon$-pseudospectrum, the other is the dynamical instability based on the linear evolution system, which is defined by the eigenvalues entering into the unstable region from the stable region under perturbations in view of pseudospectrum. The dynamical instability implies that the eigenmodes of the linear system will have exponential growth. The spectrum instability has been shown in this section, the dynamical instability will be discussed in the following section.

\section{The pseudospectrum and the transient effect}\label{pseudospectrum_and_the transients}
The properties of non-normal matrices and operators encompass various aspects, notably their ability to generate the transient effect in time-dependent dynamical systems. The reminded fact is that eigenvalues are unable to accurately describe all features of the transient effect, a natural question is that can pseudospectra provide a more accurate physical picture? The answer is unequivocally affirmative: while pseudospectra may not always furnish a precise result, they are adept at detecting and measuring the transients that eigenvalues overlook~\cite{trefethen2020spectra}. In this section, we study the correlation between the transient dynamics of a non-normal Hamiltonian $H$ which comes from Eq.(\ref{equation_of_evolution}) and its pseudospectrum.

We begin this section by writing Eq.(\ref{equation_of_evolution}) into the following linear evolution system
\begin{eqnarray}\label{linear_evolution_system}
    \frac{\mathrm{d}\mathbf{\Psi}}{\mathrm{d}v}=\mathbf{H}\mathbf{\Psi}\, ,
\end{eqnarray}
where $\mathbf{H}=\mathbf{L}_2^{-1}\mathbf{L}_1$ is a matrix acting on the function $\mathbf{\Psi}(v)$. Here, for the convenience of subsequent numerical calculation, we have already discretized $\psi(v,z)$ and the Hamiltonian $H$. Therefore, the formal solution\footnote{If $\mathbf{H}$ is a matrix or bounded operator, $e^{v\mathbf{H}}$ can be defined by a convergent power series. For an unbounded operator $\mathbf{H}$, the meaning of $e^{v\mathbf{H}}$ comes from the mathematical theory of semigroups~\cite{trefethen2020spectra}.} of the problem (\ref{linear_evolution_system}) is
\begin{eqnarray}\label{formal_solution}
    \mathbf{\Psi}(v)=e^{v\mathbf{H}}\mathbf{\Psi}(0)\, ,
\end{eqnarray}
where $\mathbf{\Psi}(0)$ is the initial condition, and we call $e^{v\mathbf{H}}$ the evolution operator in null slicing. Within the gravitational context, the evolution operator, which is adhering to the concepts proposed in~\cite{Jaramillo:2022kuv}, enables us to gain the evolution information of field, especially for that in the initial stage.

Specifically, the norm of the operator $e^{v\mathbf{H}}$ tracks the maximum rate of growth over time for the solutions $\Psi(v)$. This leads to the definition of $\lVert e^{v\mathbf{H}}\rVert_E$, i.e.,
\begin{eqnarray}\label{G(v)_definition}
    G(v)\equiv\lVert e^{v\mathbf{H}}\rVert_E=\sup_{\mathbf{\Psi}(0)\neq\mathbf{0}}\frac{\lVert e^{v\mathbf{H}}\mathbf{\Psi}(0)\rVert_E}{\lVert\mathbf{\Psi}(0)\rVert_E}\, ,
    \label{eq: Gv}
\end{eqnarray}
where the norm $\lVert\cdot\rVert_E$ is defined in Eq.(\ref{energy_norm}). The next important task is to know estimation on the size of $\lVert e^{v\mathbf{H}}\rVert_E$. It is important to convert all energy norm $\lVert\cdot\rVert_E$ related quantities into $2$-norm related quantities for the convenience of numerical calculations [see also Eq.(\ref{pseudospectrum_formula})]. For the function $G(v)$, we have 
\begin{eqnarray}\label{G(v)_computation}
    G(v)=\lVert e^{v\mathbf{H}}\rVert_E=\lVert e^{v\tilde{\mathbf{H}}}\rVert_2\, ,
\end{eqnarray}
where $\tilde{\mathbf{H}}=\mathbf{W}\mathbf{H}\mathbf{W}^{-1}$. Note that the conclusion in~\cite{trefethen_1999} has been already utilized in the second equality, and the matrix $\mathbf{W}$ is derived from the Cholesky decomposition of Gram matrix $\tilde{\mathbf{G}}^E$ (see Appendix.\ref{numerical_approach_of_pseudospectrum}). Now, computing $G(v)$ is straightforward since $2$-norm of a matrix is given by its maximum singular value so that we can do it by \textit{Mathematica}. 

In order to describe the dynamic behavior of the function $G(v)$, three abscissas are supposed to introduced, i.e., spectral abscissa $\bar{\alpha}(\mathbf{H})$, $\epsilon$-pseudospectral abscissa $\bar{\alpha}_{\epsilon}(\mathbf{H})$ and numerical abscissa $\bar{\omega}(\mathbf{H})$. It is notable that all calculations related to the three abscissas are based on $\mathbf{H}$ or $\tilde{\mathbf{H}}$, which differs from the pseudospectrum calculations of Sec.\ref{spectrum_and_pseudospectrum} by the presence of the imaginary unit $i$. Their definitions and their relationship with $G(v)$ are as follows~\cite{trefethen2020spectra,Jaramillo:2022kuv,Boyanov:2022ark}:


\begin{enumerate}
    \item The spectral abscissa $\bar{\alpha}(\mathbf{H})$ is defined by
    \begin{eqnarray}\label{spectral_abscissa}
        \bar{\alpha}(\mathbf{H})=\sup\text{Re}(\sigma(\mathbf{H}))=\sup\{\text{Re}(\lambda), \text{with}\, \lambda\in\sigma(\mathbf{H})\}\, ,
    \end{eqnarray}
where $\sigma(\mathbf{H})$ represents the spectra of $\mathbf{H}$, i.e., the QNM frequencies. In the QNM aspects, $\bar{\alpha}$ is called spectral gap which is useful in SCCC~\cite{Cardoso:2017soq}. The lower bound of $G(v)$ is controlled by the spectral abscissa,
\begin{eqnarray}\label{lower_bound}
    G(v)\ge e^{v \bar{\alpha}(\mathbf{H})}\, ,\quad \forall v\ge0\, .
\end{eqnarray}
It should be noted that even if $\bar{\alpha}<0$, i.e., the solution decays at late times, there exists the possibility of initial transient growth. 
    \item In order to estimate the maximum of the possible dynamical transient, which is dictated by the maximum growth in the norm of the evolution operator during intermediate stages of time, we are supposed to use the so-called pseudospectral abscissa. For each $\epsilon>0$, the $\epsilon$-pseudospectral abscissa $\bar{\alpha}_{\epsilon}(\mathbf{H})$ is defined by
    \begin{eqnarray}\label{pseudospectral_abscissa}
        \bar{\alpha}_{\epsilon}(\mathbf{H})=\sup\text{Re}(\sigma_{\epsilon}(\mathbf{H}))=\sup\{\text{Re}(\lambda), \text{with}\, \lambda\in\sigma_{\epsilon}(\mathbf{H})\}\, ,
    \end{eqnarray}
    where $\sigma_\epsilon(\mathbf{H})$ is the pseudospectrum of $\mathbf{H}$ (see Appendix.\ref{pseudospectrum_of_the_generalized_eigenvalue_problems}).
    In terms of the pseudospectral abscissa, the supremum of $G(v)$ can be estimated, i.e.,
    \begin{eqnarray}\label{supremum_G}
        \sup_{v\ge0} G(v)\ge \frac{\bar{\alpha}_{\epsilon}(\mathbf{H})}{\epsilon}\, ,\quad \forall \epsilon>0\, .
    \end{eqnarray}
    As a consequence of maximizing Eq.(\ref{supremum_G}) over $\epsilon>0$, this inherently results in the definitive lower bound,
    \begin{eqnarray}
        \sup_{v\ge0} G(v)\ge \mathcal{K}(\mathbf{H})\, .
    \end{eqnarray}
    The constant $\mathcal{K}(\mathbf{H})$ is called the Kreiss constant of $\mathbf{H}$, which is defined by $\sup_{\epsilon > 0} \bar{\alpha}_{\epsilon}/\epsilon$. In Fig.\ref{fig:KreissConstant}, we depict its tendency  with benchmark parameters as $\epsilon$ increases, the $\epsilon$-pseudospectral absicissa $\bar{\alpha}_{\epsilon}$ is calculated by the criss-cross algorithm~\cite{trefethen2020spectra,10.1093/imanum/23.3.359}. Obviously, the Kreiss constant in the set of benchmark parameters equals to $1$.
    \begin{figure}[htbp]
        \centering
        \includegraphics{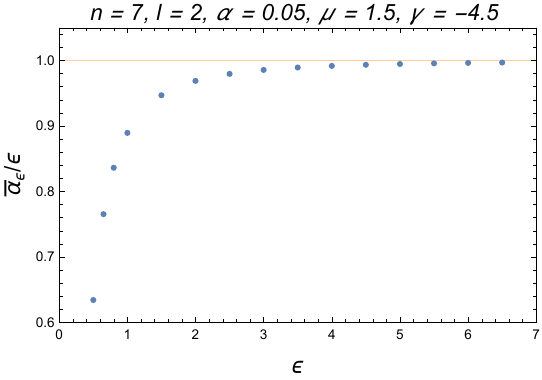}
        \caption{The tendency of $\bar{\alpha}_{\epsilon}/\epsilon$ varies with $\epsilon$ from $0.5$ to $6.5$, and $\delta\epsilon=0.5$. The blue points are discrete data, which are calculated with the benchmark parameters, and the horizontal orange line corresponds to $\bar{\alpha}_{\epsilon}/\epsilon = 1$. The resulotion $N = 100$ and $100$ digits of precision are used.}
        \label{fig:KreissConstant}
    \end{figure}
    
    \item The numerical abscissa $\bar{\omega}(\mathbf{H})$ can be obtained by 
    \begin{eqnarray}\label{numerical_abscissa}
        \bar{\omega}(\mathbf{H})=\sup\sigma\Big(\frac{1}{2}(\mathbf{H} + \mathbf{H}^{\dagger})\Big)\, .
    \end{eqnarray}
    Note that the symbol $\dagger$ means the adjoint of an operator (matrix). The discrete version $\mathbf{H}^{\dagger}$ can be obtained from
    \begin{eqnarray}
        \mathbf{H}^{\dagger}=(\tilde{\mathbf{G}}^E)^{-1}\cdot\mathbf{H}^{\star}\cdot\tilde{\mathbf{G}}^E\, ,
    \end{eqnarray}
    where the symbol $\star$ stands for the conjugate transpose. The numerical abscissa in Eq.(\ref{numerical_abscissa}) characterizes the initial slope of the norm of the evolution operator $e^{v\mathbf{H}}$, which means that
    \begin{eqnarray} \label{the relation between numerical abscissa and G(v)}
        \bar{\omega}(\mathbf{H})=\frac{\mathrm{d}}{\mathrm{d}v}\lVert e^{v\mathbf{H}}\rVert_E\Big|_{v=0}\, .
    \end{eqnarray}
    Therefore, if the value $\bar{\omega}(\mathbf{H})>0$, there will exist an initial growth in the evolution. The upper bound of $G(v)$ is controlled by the numerical abscissa,
    \begin{eqnarray}\label{upper_bound}
        G(v)\le e^{v \bar{\omega}(\mathbf{H})}\, ,\quad \forall v\ge0\, .
    \end{eqnarray}
    If $\bar{\omega}=0$, $G(v)$ fails to exhibit growth at initial stage, and diminishes as $v$ increases, which signifies the Kreiss constant must inherently equal unity, i.e., $\mathcal{K}=1$~\cite{trefethen2020spectra}. But the reverse is not necessarily true. The model we are studying is a counterexample where its numerical abscissa is larger than $0$ but the Kreiss constant is still $1$. The overall numerical abscissa are shown in Fig.\ref{fig:numerical abscissa} varying with different parameters.

    \begin{figure}[htbp]
    \centering
    \includegraphics[width = 0.3\textwidth]{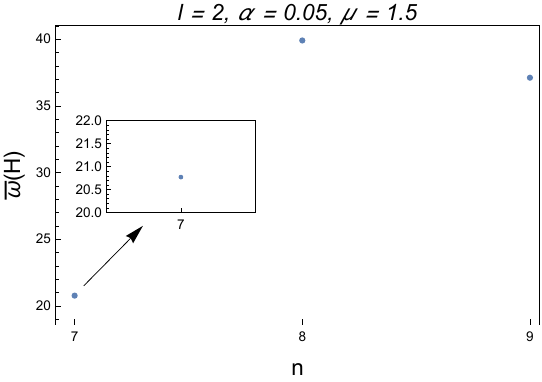}
    \includegraphics[width = 0.31\textwidth]{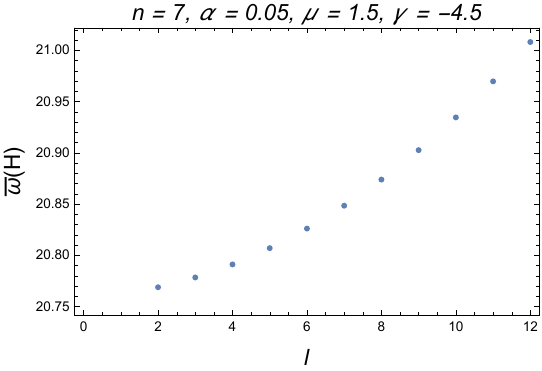}
    \includegraphics[width = 0.3\textwidth]{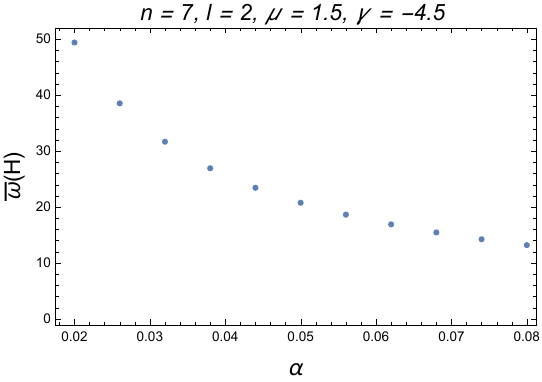}
    \includegraphics[width = 0.31\textwidth]{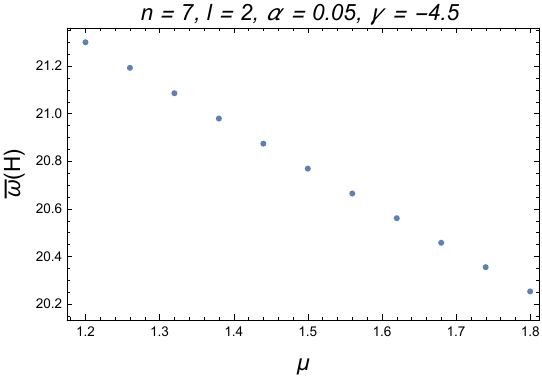}
    \includegraphics[width = 0.3\textwidth]{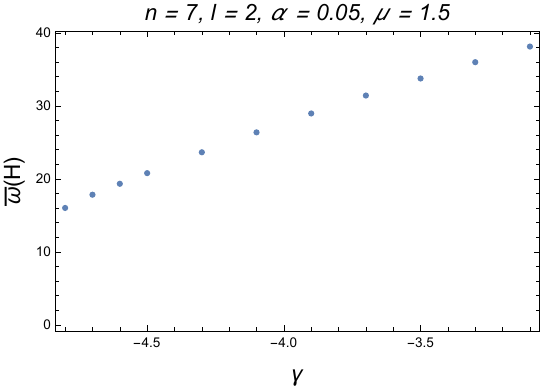}
    \caption{The numerical abscissa varies with different parameters $n$, $\alpha$, $l$, $\mu$ and $\gamma$ with resolution $N=100$ and $100$ digits of precision. The numerical abscissa corresponding to the benchmark parameters approximately equals to $20.8$, which has been amplified in the first panel labeled by the black arrow. }
    \label{fig:numerical abscissa}
\end{figure}
\end{enumerate}

\subsection{The distance to the dynamical instability}
As mentioned earlier, there are two types of instability: spectrum instability and dynamical instability. Spectrum instability is characterized by the open structure of the pseudospectrum, as discussed in Section \ref{spectrum_and_pseudospectrum}. Dynamical instability, on the other hand, refers to the behavior of a linear system exhibiting exponential growth. The behavior of a linear system is well-known to be controlled by the spectral abscissa, as shown in Eq. (\ref{lower_bound}). If the spectrum abscissa is greater than zero, exponential growth occurs. In other words, a linear system is considered dynamically stable if the spectrum lies entirely within the open left half-plane. Conversely, if the spectrum extends into the right half-plane, the system is considered dynamically unstable. Actually, these two types of instability are closely related. Spectrum instability is responsible for the significant shift of the spectrum from its initial state when subjected to external perturbations. This shift can even result in the spectrum migrating into the so-called dynamically unstable region.

\begin{figure}[htbp]
    \centering
    \includegraphics[width = 0.5\textwidth]{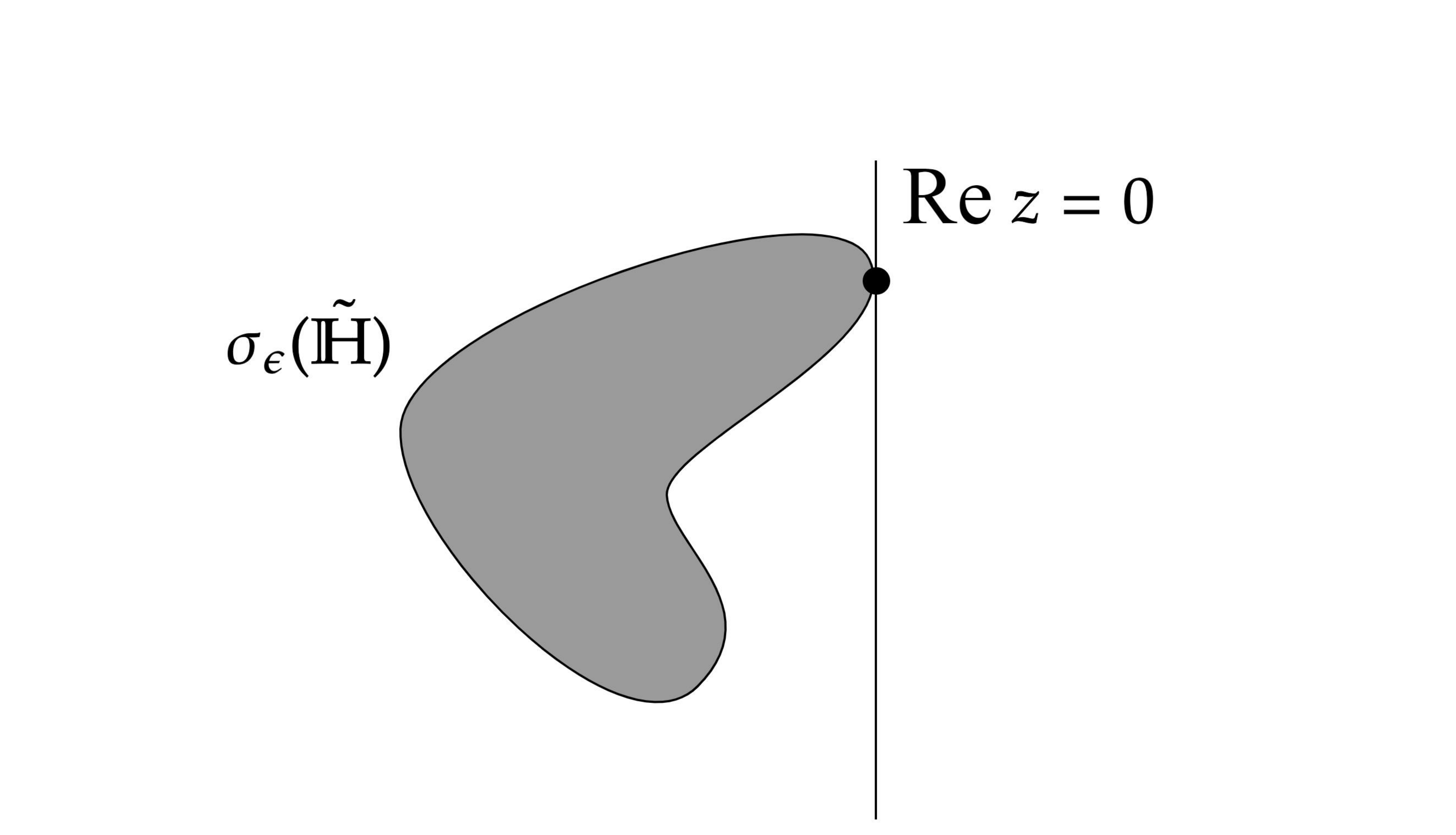}
    \caption{The instability threshold for $\tilde{\mathbf{H}}$ corresponds to the minimal value of $\epsilon$ at which the boundary of set $\sigma_{\epsilon}(\tilde{\mathbf{H}})$ intersects the imaginary axis.}
    \label{fig:distance_to_instability}
\end{figure}

Naturally, how much external perturbation would lead to the spectrum migrating to the dynamically unstable region? Although this is a very physical problem, we can treat it as a mathematical problem. Therefore, it is natural to investigate the distance to dynamical instability of $\tilde{\mathbf{H}}=\mathbf{W}\mathbf{H}\mathbf{W}^{-1}$, which controls the evolution direction of the entire system. Its definition is given by~\cite{trefethen2020spectra}
\begin{eqnarray}
    \min \Big\{\lVert\mathbf{E}\rVert_2: \tilde{\mathbf{H}}+\mathbf{E}\, \text{is unstable} \Big\} \, .
    \label{definition of unstable}
\end{eqnarray}
The interpretation of Eq.(\ref{definition of unstable}) in terms of pseudospectra is straightforward: what is the smallest $\epsilon$ for which the boundary of $\sigma_{\epsilon}(\tilde{\mathbf{H}})$ touches the imaginary axis? In Fig.\ref{fig:distance_to_instability}, the shaded area represents all possible ranges of the spectra after being disturbed by an external intensity of $\epsilon$. As $\epsilon$ increases, we always see that there exists such $\epsilon$ that the boundary of $\sigma_{\epsilon}(\tilde{\mathbf{H}})$ is just tangent to the imaginary axis. At this time, we call such $\epsilon$ is the (complex) stability radius of $\tilde{\mathbf{H}}$. Given a linear system whose Hamiltonian is given by $\tilde{\mathbf{H}}$, then the theorem~\cite{trefethen2020spectra} below provides us with a method for calculating the radius of (complex) stability.

\begin{theorem}
    The $2$-norm complex stability  radius of a matrix $\tilde{\mathbf{H}}$ is equal to 
    \begin{eqnarray}\label{complex_stability_radii}
        \Bigg(\sup_{\text{Re}z=0}s_1\Big(\mathbf{R}(z)\Big)\Bigg)^{-1}\, ,
    \end{eqnarray}
    where $s_1$ denotes the largest singular value and $\mathbf{R}(z)=(z-\tilde{\mathbf{H}})^{-1}$.
\end{theorem}

\begin{figure}[htbp]
    \centering
    \includegraphics[width = 0.3\textwidth]{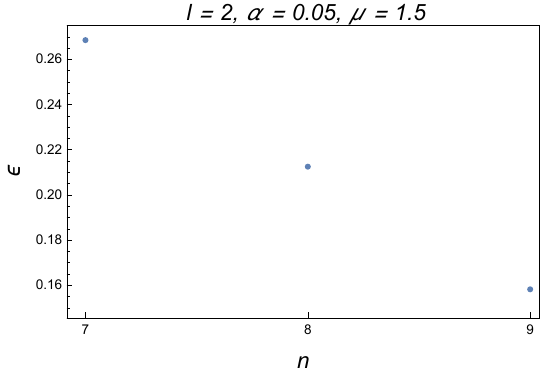}
     \includegraphics[width = 0.3\textwidth]{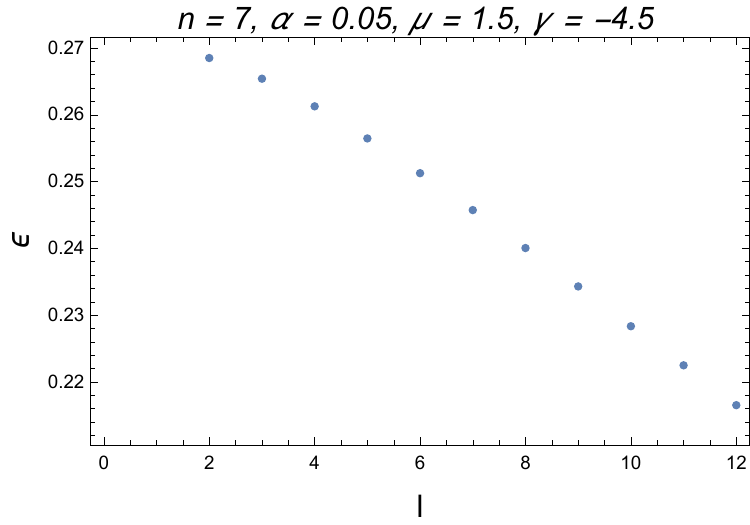}
    \includegraphics[width = 0.3\textwidth]{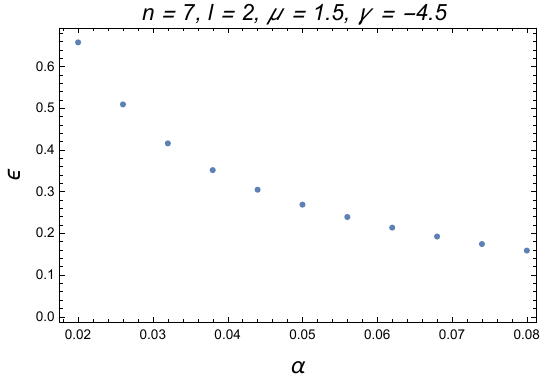}
    \includegraphics[width = 0.3\textwidth]{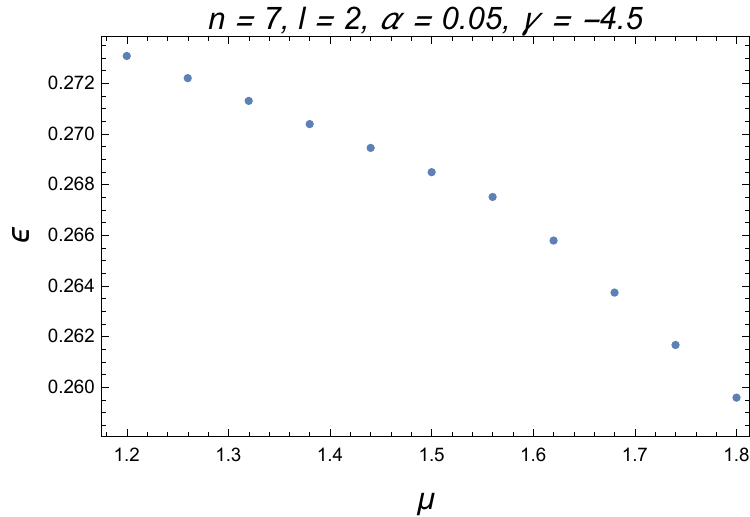}
    \includegraphics[width = 0.3\textwidth]{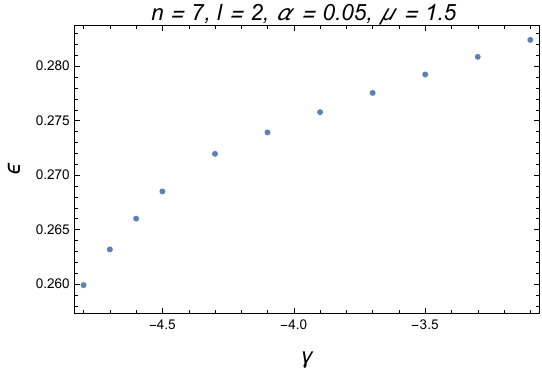}
    \caption{The complex stability radius $\epsilon$ changes with different parameters $n$, $l$, $\alpha$, $\mu$ and $\gamma$ with resolution $N=100$ and $100$ digits of precision. The each parameter has eleven points in the second panel from the fifth panel while the range of spacetime dimension $n$ is limited into $n= 7,8,9$ in the first panel.}
    \label{fig:Complex Radius}
\end{figure}

In this section, we use Eq.(\ref{complex_stability_radii}) to calculate the complex stability radius and present some typical results varying with different parameters in Fig.\ref{fig:Complex Radius}. By examining the five panels, we can observe certain patterns. In the second to fourth panels, we find that the stability radius decreases with increasing $l$, $\alpha$, $\mu$. However, in contrast, the stability radius increases with increasing $\gamma$. In the first panel, the stability radius for two different spacetime dimensions $n=7$ and $n=8$ can be compared due to the same $\gamma$, and the former is more dynamically stable than the latter. For the case of $n=9$, since we choose $\gamma=-6.5$ as illustrated before, it can not be compared with other cases. The complex stability radius provides a quantitative measure of dynamical instability of the linear system. Based on the results, we can infer that increasing $n,l,\alpha,\mu$ will destabilize the dynamical system, while increasing $\gamma$ will stabilize it, since the larger the stability radius, the greater the perturbation required to complete the transition of the system from dynamic stability to dynamic instability.

\begin{figure}[htbp]
    \centering
    \includegraphics[width = 0.3\textwidth]{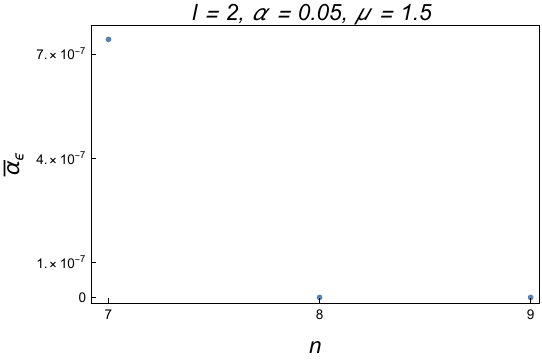}
    \includegraphics[width = 0.3\textwidth]{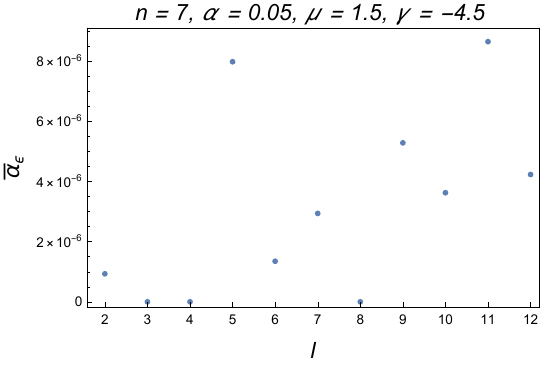}
    \includegraphics[width = 0.3\textwidth]{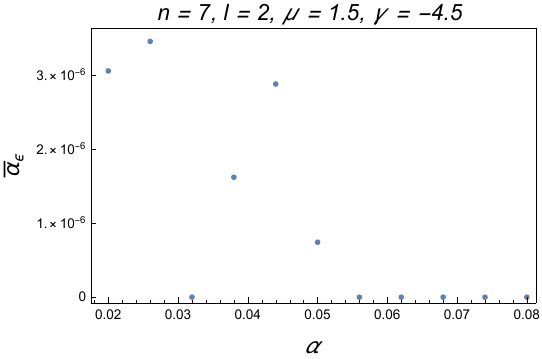}
    \includegraphics[width = 0.3\textwidth]{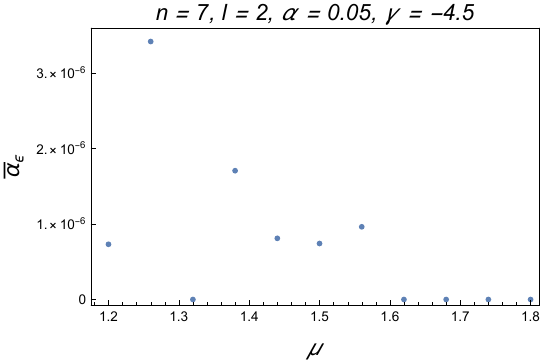}
    \includegraphics[width = 0.3\textwidth]{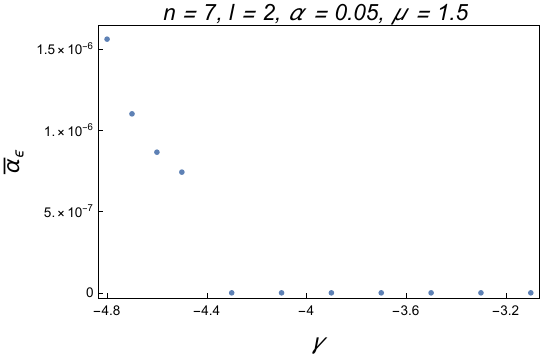}
    \caption{The pseudospectral abscissa $\bar{\alpha}_{\epsilon}$ changes with different parameters $n$, $l$, $\alpha$, $\mu$ and $\gamma$ with resolution $N=100$ and $100$ digits of precision, in which $\epsilon$ is selected as the complex stability radius.}
    \label{fig:Spectrum abscissa depended on the complex radius}
\end{figure}

In addition, the stability radius can also be determined by the pseudospectral abscissa $\bar{\alpha}_{\epsilon}(\tilde{\mathbf{H}})=0$ by definition. The criss-cross algorithm is a commonly used approach to calculate the pseudospectral abscissa. More details about the criss-cross algorithm can be found in references~\cite{10.1093/imanum/23.3.359, trefethen2020spectra}. In general, we can employ the bisection method to iteratively approach the desired (complex) stability radius. On the other hand, this also can serve as a method to verify the reliability of the stability radius results obtained from Eq.(\ref{complex_stability_radii}). To evaluate its accuracy, we can calculate the pseudospectral abscissas where $\epsilon$ is given by the stability radius, thereby examining the errors. The pseudospectral abscissas with parameters that are same with calculations in stability radius are shown in Fig.\ref{fig:Spectrum abscissa depended on the complex radius}. From this figure, we observe that the pseudospectral abscissas are controlled within the order of magnitude of $10^{-6}$, indicating relatively small errors, that provides further validation of the stability radius results.

\subsection{The transient effect and the waveform}

In this subsection, we will investigate the transient effect and the waveform of Maeda-Dadhich black hole in EGB gravity theory. The transient effect is characterized by the growth of the norm of the evolution operator $e^{v\mathbf{H}}$, where the norm is given by the energy norm [see Eq.(\ref{energy_norm})]. Therefore, it is crucial to have a comprehensive understanding of the evolution of the system described in Eq.(\ref{formal_solution}) across different stages. The evolution governed by the operator $e^{v\mathbf{H}}$ can be divided into three distinct phases in view of its norm. Firstly, the spectral abscissa $\bar{\alpha}(\mathbf{H})$ dominates the asymptotic behavior of the linear system in the limit of late time $v\to\infty$. Secondly, the numerical abscissa $\bar{\omega}(\mathbf{H})$ controls the initial growth as $v\to 0$ [see Eq.(\ref{upper_bound})]. The $\epsilon$-pseudospectrum concept plays a key role in bridging the late-time dynamics governed by $\bar{\alpha}(\mathbf{H})$ and the early-time behavior dictated by $\bar{\omega}(\mathbf{H})$. It provides a means to control the growth at intermediate times using the pseudospectral abscissa $\bar{\alpha}_{\epsilon}(\mathbf{H})$ [see Eq.(\ref{supremum_G})].

\begin{figure}[htbp]
    \centering
    \includegraphics[width = 0.5\textwidth]{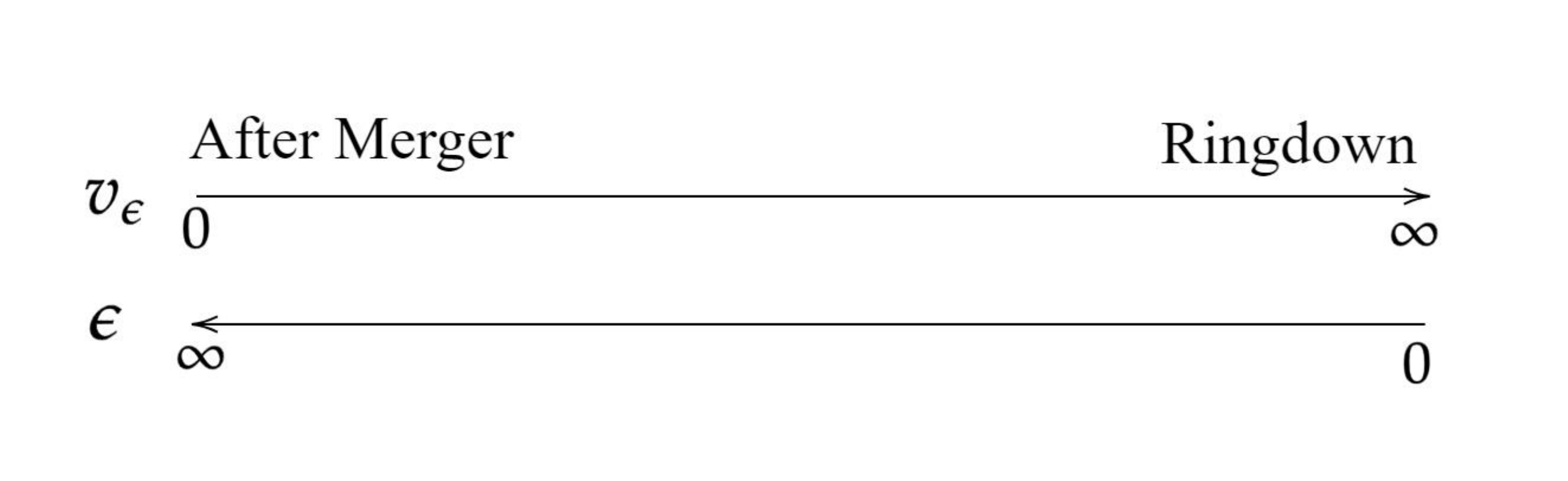}
    \caption{The sketch map that describe the relationship between $v_{\epsilon}$ and $\epsilon$.}
    \label{fig:relationship}
\end{figure}

On the other hand, the three abscissas have the relationship with perturbation size $\epsilon$. In the pioneering work~\cite{Jaramillo:2022kuv}, the author study the BBH merger waveform as a transient phenomenon under the (strong) hypothesis of effective BBH non-normal linear dynamics, and illustrate that the (late) inspiral phase corresponding to the limit $\epsilon \to \infty$, the final ringdown characterised by $\epsilon \to 0$ and the actual merger by intermediate values of $\epsilon$. However, within GWs context in view of pseudospectrum, we consider that the limit $\epsilon \to \infty$ corresponds the phase after the merger is completed, based on the following reasons. Firstly, pseudospectrum describes the properties of operator $H$ in Eq.(\ref{linear_evolution_system}) in linear evolution system, which origins from the master equation (\ref{Schrodinger_equation}). And the master equation describes the properties of a single black hole distorted by some perturbations, which doesn't involve how the binary black holes evolve specifically. Thus, the limitation $\epsilon \to \infty$ should correspond to the phase after the merger is completed, where a single event horizon has been formed, at that moment the black hole is at intense distortion. Then naturally, the late time dominated by ringdown phase corresponds to the limitation $\epsilon \to 0$, whose evolution behaviour is characterised by QNMs, i.e. spectral abscissa. We will use a schematic diagram to illustrate this process, see Fig.\ref{fig:relationship}.

However, the three types of abscissas provide rough estimates of $G(v)$. In order to obtain a more accurate grasp of the behavior of $G(v)$, we can utilize equation (\ref{G(v)_computation}) to establish the relationship between $G(v)$ and time. It is important to clarify that there is another formulation presented in an other work~\cite{Carballo:2024kbk}. In that work, they define $G(\tau)$ using the norm-squared, resulting in $G^2(v)$ in our work. However, despite the difference in notation, both formulations capture the same physical meaning as they utilize the energy norm. It is also worth noting that $G(v)$ represents the maximum possible energy growth of the field at time $v$.

\begin{figure}[htbp]
    \centering
    \includegraphics[width = 0.5\textwidth]{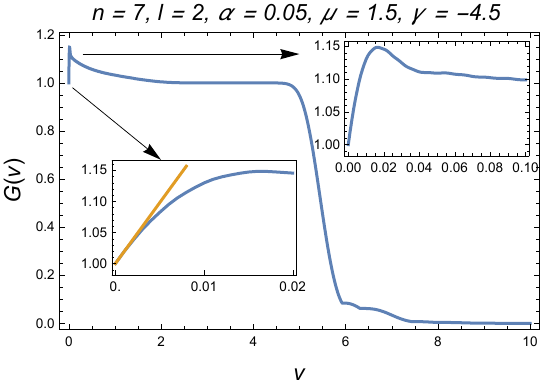}
    \caption{The relationship between the norm $G(v)$ of the evolution operator $e^{v\mathbf{H}}$ and the time $v$ within the benchmark parameters.}
    \label{fig:TransientGrowth}
\end{figure}

As a typical example, we depict the variation of $G(v)$ with respect to $v$ using the benchmark parameters, as shown in Fig.\ref{fig:TransientGrowth}. The blue line represents the function $G(v)$. In the lower left amplified subgraph, the orange line represents the tangent line of $G(v)$ at $v=0$, and its slope corresponds to the numerical abscissa (see Fig.\ref{fig:numerical abscissa}). In the initial stage, there is a noticeable growth of $G(v)$ controlled by the numerical abscissa, with values larger than $1$, as indicated in the two amplified subgraphs. This behavior is fundamentally different from the hyperboloidal case presented in~\cite{Carballo:2024kbk}, where $G(\tau)$ is less than or equal to $1$. The reason for this difference lies in the fact that in the hyperboloidal case, $\partial_{\tau}E(\tau)$ can be proven to be less than or equal to zero. However, in our case, it is difficult to determine the positivity or negativity of $\partial_{v}E[\varphi]$ [see Eq.(\ref{partial energy v})]. As a result, there are no such limitations on the behavior of $G(v)$. In the intermediate stage, we observe a platform in the $G(v)$ curve. These two properties of $G(v)$ strongly suggest the existence of the transient effect. The reason behind these properties is that although the presence of an event horizon ensures energy overflow, the eigenmodes (i.e., QNMs) are not completely orthogonal due to the non-normality of the operator $H$. This lack of orthogonality allows for energy exchange or transfer between non-orthogonal modes, resulting in the transient effect, even though the system remains at a linear level. In the late stage, starting at $v_{\star}\approx4.7$, we observe a decline in $G(v)$, which is dominated by the spectral abscissa. 

\begin{figure}[htbp]
            \centering
            \includegraphics[width=0.3\linewidth]{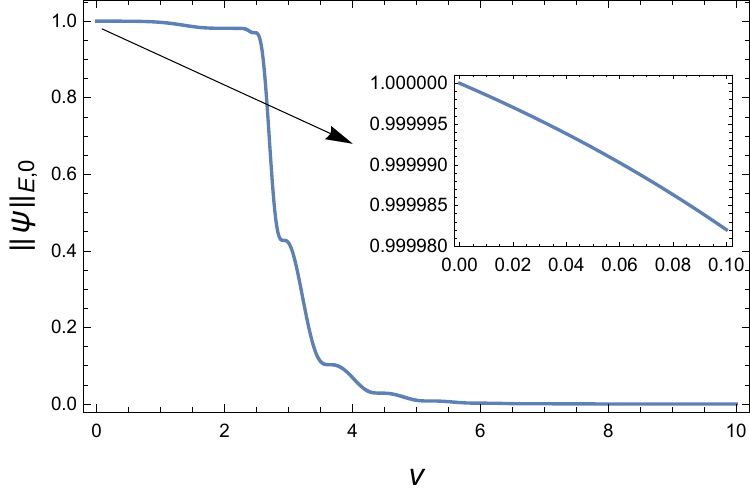}
            \includegraphics[width=0.3\linewidth]{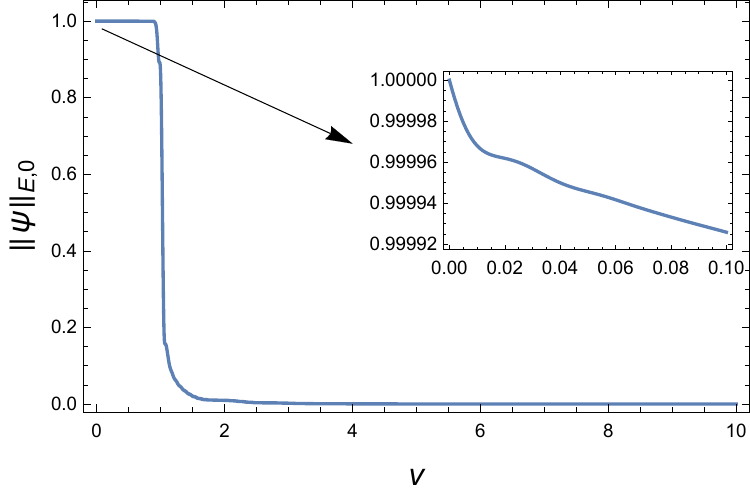}
             \includegraphics[width=0.3\linewidth]{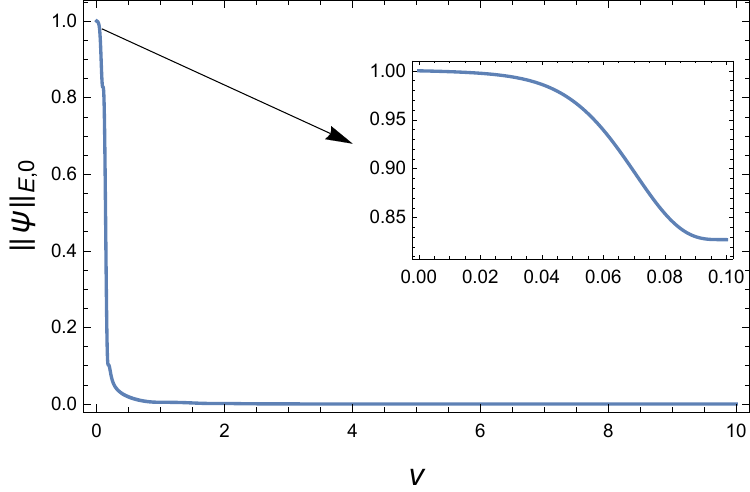}
            \includegraphics[width=0.3\linewidth]{EnergyNormPart_1.pdf}
             \includegraphics[width=0.3\linewidth]{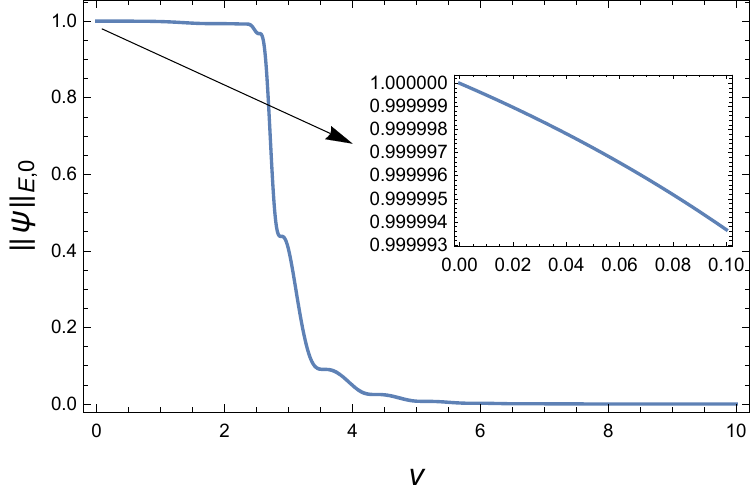}
             \includegraphics[width=0.3\linewidth]{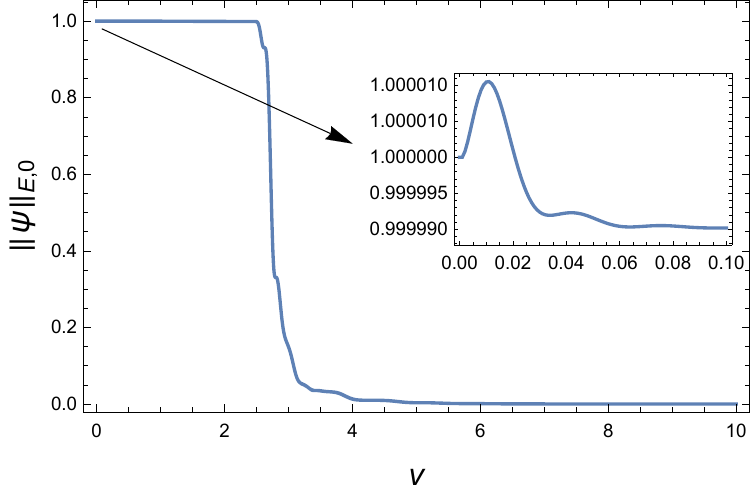}
            \caption{The results of evolution of the normalized energy norm of a wave packet. The parameters of initial positions and widths are $z_{0} = 0.95, 0.5, 0.05$ corresponding to the panels from left to right in the first row  while $\lambda = \sqrt{10}/100$, and $\lambda = \sqrt{10}/100, \sqrt{5}/100, \sqrt{1}/100$ corresponding to the panels from left to right in the second row  while $z_{0} = 0.95$, respectively. The first colunm panels are same within $\lambda=\sqrt{10}/100$ and $z_0=0.95$.}
            \label{fig:energy norm}
\end{figure}

\begin{figure}[htbp]
    \centering
    \includegraphics[width=0.3\textwidth]{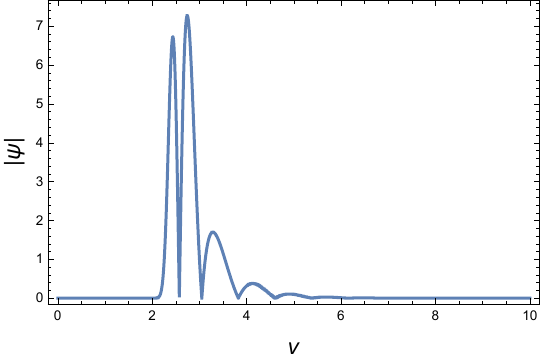}
    \includegraphics[width=0.3\textwidth]{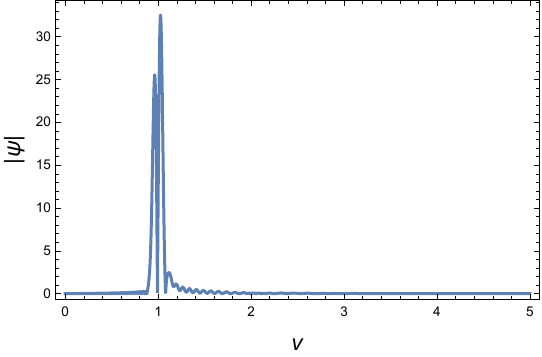}
    \includegraphics[width=0.3\textwidth]{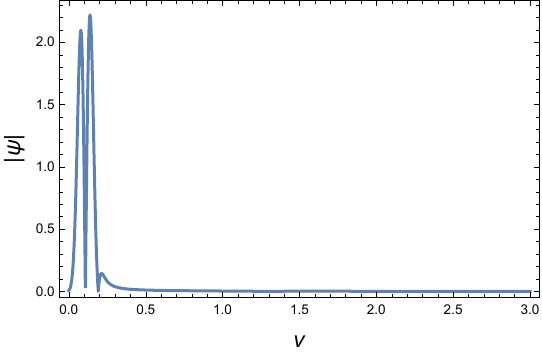}
    \includegraphics[width=0.3\textwidth]{IniWavePos_1.pdf}    
    \includegraphics[width=0.3\textwidth]{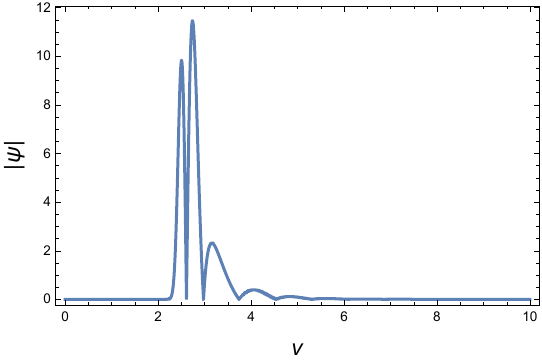}
    \includegraphics[width=0.3\textwidth]{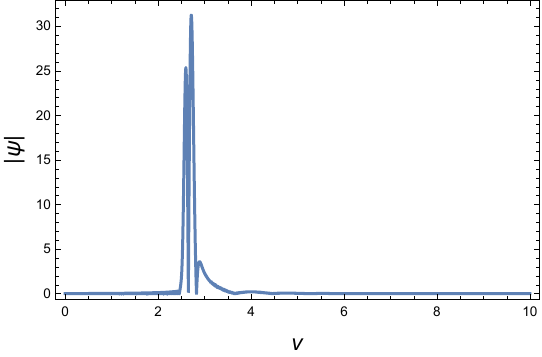}
    \includegraphics[width=0.3\textwidth]{IniWavePos_1.pdf}
    \includegraphics[width=0.3\textwidth]{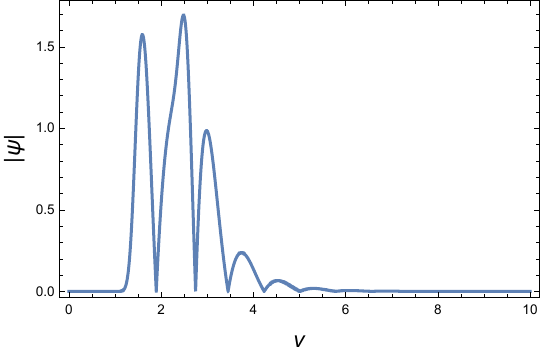}
    \includegraphics[width=0.3\textwidth]{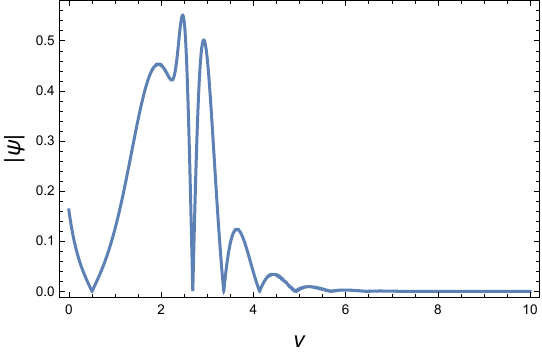}
    \caption{The nine panels are the waveform calculated with Eq.(\ref{linear_evolution_system}) by using time-symmetric numerical method (see Appendix.\ref{numerical_method}). The initial condition is set as a Gaussian wave packet with amplitude $h=1$. The first and second row panels are the results of waveform observed at null infinity $z_{\text{obs}}=0$ varying with the initial positions and widths of wave packet, where parameters are $z_{0}= 0.95, 0.5, 0.05$ and $\lambda = \sqrt{10}/100, \sqrt{5}/100, \sqrt{1}/100$ from left to right, respectively. In third row panels, we change the observation positions, corresponding to $z_{\text{obs}} = 0, 0.5, 0.96$ from left to right, respectively. Therefore, the first column panels are the same waveform observed at null infinity $z_{\text{obs}}=0$, where parameters of initial Gaussian wave packet are $\lambda = \sqrt{10}/100$ and $z_{0}=0.95$.}
    \label{fig:waveform}
\end{figure}

In order to illustrate the relation between the transient growth of energy norm and the analysis of $G(v)$, we depict the evolution of normalised energy norm as it varies with different initial positions and initial widths of the wave packet, while keeping the amplitude unchanged. These results are shown in Fig.\ref{fig:energy norm} and the normalised energy norm can be computed by using
\begin{eqnarray} \label{eq; normalised energy norm evolution}
    \lVert \mathbf{\Psi} (v)\rVert_{E,0} = \frac{\sqrt{\mathbf{\Psi^{\star}}(v) \cdot \mathbf{\Tilde{G}}^{E} \cdot \mathbf{\Psi}(v)}}{\sqrt{\mathbf{\Psi^{\star}}(0) \cdot \mathbf{\Tilde{G}}^{E} \cdot \mathbf{\Psi}(0)}} \, .
\end{eqnarray}
Additionally, in traditional gravitational wave observations, there is a focus on how the waveform $|\Psi|$ changes over time. The waveform typically refers to the variation of the field or its absolute value with respect to time. Motivated by this, we also depict the waveform as it varies with same parameters used in depicting the evolution of energy norm, and we add the observation position as a new variation. The results are shown in Fig.\ref{fig:waveform}. The initial condition of both cases $\psi(0,z)$ is set as a Gaussian wave packet, given by:
\begin{eqnarray}
    \psi(0,z)=h\exp\Big[-\frac{(z-z_0)^2}{2\lambda^2}\Big]\, ,
    \label{eq: initial wave packet}
\end{eqnarray}
where $z_0$ is the initial position, $\lambda$ is the width and $h=1$ is the amplitude, respectively. After spatial discretization, we solve the linear evolution Eq.(\ref{linear_evolution_system}) by using the time-symmetric numerical method~\cite{markakis2019timesymmetry,OBoyle:2022yhp,Markakis:2023pfh} instead of the traditional Runge-Kutta method, see more details in Appendix.\ref{numerical_method}.

The results illustrated in Fig.\ref{fig:energy norm} indicate its evolution behavior. The initial wave packet is structured in accordance with Eq.(\ref{eq: initial wave packet}). The parameters governing the initial positions and widths are $z_{0}=0.95,0.5,0.05$ corresponding to the panels from left to right in the first row while $\lambda = \sqrt{10}/100$, and $\lambda=\sqrt{10}/100,\sqrt{5}/100,\sqrt{1}/100$ corresponding to the panels from left to right in the second row while $z_{0} = 0.95$, respectively. In the last panel, where the initial wave packet parameters are set to $z_{0} = 0.95$ and $\lambda = \sqrt{1}/100$, a marginal initial transient growth of the energy norm is observed in the magnified subplot. This growth, although slight, contrasts with the absence of similar growth in the other panels. These findings align with the analysis of $G(v)$ conducted using Eq.(\ref{eq: Gv}) since $G(v)$ characterizes the maximum potential initial transient growth of the energy norm of the field. The outcome displayed in the last panel of Fig.\ref{fig:energy norm} represents one potential manifestation of transient growth within this context.

In Fig.\ref{fig:waveform}, the first column panels depict the same waveform, where the initial conditions are given by $z_{0} = 0.95$ and $\lambda = \sqrt{10}/100$. The waveform is observed at null infinity, $z_{\text{obs}}=0$. It is observed that after an amplification in amplitude, which begins at approximately $v\approx 2.2$, the wave packet starts to exponentially decay. The first row panels show the results when varying the initial positions of the Gaussian wave packet. The three panels from left to right correspond to $z_{0}=0.95$, $z_{0}=0.5$, and $z_{0}=0.05$. The qualitative characteristics of the waveform in the second and third figures are consistent with those in the first figure, but there are two notable features. Firstly, the closer the initial position of the wave packet is to the observation position, the earlier the wave packet is observed. Secondly, when the initial wave packet is located at $z=0$, the maximum amplitude of the wave packet is observed at null infinity. The second row panels show the results when varying the initial width of the Gaussian wave packet. The three panels from left to right correspond to $\lambda = \sqrt{10}/100$, $\lambda = \sqrt{5}/100$, and $\lambda =\sqrt{1}/100$. These results indicate that the width of the Gaussian wave packet affects the duration of the wave packet at the observation position, which is consistent with~\cite{Baibhav:2023clw}. The third row panels show the results when varying the observation positions. The three panels from left to right correspond to $z_{\text{obs}}= 0$, $z_{\text{obs}}= 0.5$, and $z_{\text{obs}}= 0.96$. From these panels, we observe that the closer the observation position is to null infinity, the greater the amplitude of the waveform obtained.

It is important to note that the moment when the exponential decay begins for $G(v)$ does not necessarily correspond to the moment when the exponential decay begins for the waveform, which is basic question in BBH dynamics that when is ringdown excited. The waveform depends on the spatial observation position, and further discussions about the excitation of the ringdown in terms of the magnitude $|\Psi|$ at fixed observation positions can be found in~\cite{Baibhav:2023clw}, while $G(v)$ does not since it has integrated out the spatial part. However, the existence of transient growth in $G(v)$ does provide some insights about the waveform observed at a fixed spatial position. It can at least be concluded that the existence of transient growth allows for the amplitude of the wave received at infinity to be larger than the initially given amplitude. The initial width and position of the wave packet do not change the fact that the amplitude is amplified at null infinity.


\section{conclusions and discussion}\label{conclusions_and_discussion} 
In this paper, we investigate the pseudospectrum and the transient effect of the tensor perturbation for Maeda-Dadhich black holes in EGB gravity theory. The topology of the so-called Maeda-Dadhich black holes with dimension $n$ is the product of a usual four-dimensional spacetime with a negative constant curvature space with dimension $n-4$. Upon examining the asymptotic expansion of the metric function, it is observed that the behavior of the metric function resembles the one of the Reissner-Nordstr\"{o}m anti–de Sitter spacetime, albeit in the absence of the Maxwell field.

With regard to technical details, we have devised the pseudospectrum computation by using ingoing Eddington-Finkelstein coordinates, which is also used in~\cite{Cownden:2023dam,Boyanov:2023qqf}. The equation governing the perturbations undergoes a reduction to a first-order temporal form, with the time derivative appearing in a mixed derivative term~[see Eq.(\ref{equation_of_evolution})]. This transition naturally prompts a reformulation of the pseudospectrum as a generalized eigenvalue problem. The adoption of these coordinates, coupled with an appropriate rescaling of the field, not only drastically simplifies the problem and facilitates the imposition of boundary conditions, but also remarkably decreases the dimension of the related matrices (Within the same resolution $N$, the matrix dimension in the hyperboloidal setting is twice that in the null setting.), thereby lessening the computational demands of the numerical implementation. Based on the analysis of the convergence of the hyperbolidal slicing and the null slicing for the pseudospectra in~\cite{Boyanov:2023qqf}, we are convinced that the calculation of pseudospectra under the null slicing is appropriate.

The spectrum and pseudospectrum of the model we studied is given in Sec.\ref{spectrum_and_pseudospectrum}. We have depicted the behaviour of the fundamental mode with parameters changing in Fig.\ref{fig:fundamentalMOde changes with parameters}. The decay time of fundamental mode is decreasing as $\mu$ and $\gamma$ increase, but increasing as $\alpha$ and $l$ increase, while the fundamental mode is purely imaginary when $n=8$ and $n=9$. In the Fig.\ref{fig:pseudospectrum}, the ``inverted bridge'' structure of pseudospectrum is displayed. Such type of structure reflects the spectral instability.

In order to further study the pseudospectrum and the transient effect of our model, where the transient effect is described by $G(v)$. Moreover, three kinds of abscissas, i.e., the spectral abscissa $\bar{\alpha}$, the $\epsilon$-pseudospectral abscissa $\bar{\alpha}_{\epsilon}$ and the numerical abscissa $\bar{\omega}$, are used to describe the dynamics of $G(v)$. First, the pseudospectral abscissa is used to define the distance to the dynamical instability, which is also called the (complex) stability radius. The larger stability radius represents that making the spectral abscissa into dynamical unstable region requires larger external perturbation energy. We calculate such stability radii within different parameters, increasing $n$, $l$, $\alpha$ and $\mu$ reduces the dynamical stability, but increasing $\gamma$ enhances the dynamical stability. It is notable that there is no guaranteed relationship between the decay time and the stability radius. 

More importantly, we firstly report the transient effect in null slicing with ingoing EF coordinates.
In addition, the numerical abscissas are larger than $0$ that confirms the the behaviour of $G(v)$ at initial stage, which is consistent with the analysis of $E[\varphi]$, since there is a relationship between them that the numerical abscissa equals to the time derivative of $G(v)$ at $v=0$ [see Eq.(\ref{the relation between numerical abscissa and G(v)})]. The case is essentially different with the hyperboloidal case~\cite{Carballo:2024kbk}. Then we explain the relationship between the transient effect and the evolution of normalised energy norm, as well as waveform, which are solved by Eq.(\ref{eq; normalised energy norm evolution}) and Eq.(\ref{linear_evolution_system}) with a initial Gaussian wave through the numerical method. The initial growth of normalised energy norm shown in last panel of Fig.\ref{fig:energy norm} are consistent with the analysis of $G(v)$. However, the analyses conducted on $G(v)$ and the evolution of the energy norm of a field primarily merely hint at the potential for dynamical instability through the breakdown of perturbation theory. To conclusively demonstrate dynamical instability, a nonlinear analysis would be imperative. This deeper dive into nonlinear dynamics is crucial for substantiating the presence and implications of instability within the system. Additionally, the transient effect enables the amplitude of the wave received at null infinity to exceed the initially given amplitude. Although the wide of Guassian wave packet will suppress the duration of waveform observed, the initial conditions of the wave packet do not change the qualitative fact that the amplitude is amplified at null infinity.

There are some things that can be further explored. Here, we list some investigations that can be done in the future in terms of pseudospectrum for the gravitational physics.
\begin{enumerate}
    \item In the general relativity, the positive definiteness of the effective potential outside the event horizon guarantees the positive definiteness of the Gram matrix $\mathbf{G}$, so that the Cholesky decomposition can always be done. However, in some modified gravity theories, this is not always the case. The discussion of pseudospectra in cases with the large potential well occurring will be done in the future.

    \item In the definition of (complex) pseudospectrum, the elements of the perturbation matrix $\mathbf{E}$ are generally complex. However, considering the physical reality, the elements of matrix $\mathbf{E}$ should be real, which imposes an important restriction on pseudospectrum. In this case, the pseudospectrum is referred to as the real pseudospectrum, and its structure differs significantly from that of the complex pseudospectrum mentioned earlier. Then, extending to a more general case, we have the following definition of structured $\epsilon$-pseudospectrum~\cite{trefethen2020spectra}:
    \begin{eqnarray}
        \bigcup_{\mathbf{E}\, \text{structured}\, , \lVert\mathbf{E}\rVert<\epsilon}\sigma(\mathbf{H}+\mathbf{E})\, .
    \end{eqnarray}
    More work regarding structured pseudospectra can be referenced in~\cite{GRAILLAT200668,noschese2016approximatedstructuredpseudospectra}. In gravitational physics, exploring the differences between complex pseudospectra and structured pseudospectra is worthwhile, and such work will be completed in the future.
\end{enumerate}
\section*{Acknowledgement}
We are grateful to Li-Ming Cao, Yu-Sen Zhou, Long-Yue Li and Xia-Yuan Liu for helpful discussions. This work is supported in part by the National Key Research and Development Program of China Grant No.2020YFC2201501, in part by the National Natural Science Foundation of China under Grant No.12075297 and No.12235019.

	
\appendix

\section{The pseudospectrum of the generalized eigenvalue problems}\label{pseudospectrum_of_the_generalized_eigenvalue_problems}
In this appendix, we will follow the monograph~\cite{trefethen2020spectra} and give some definitions of the pseudospectrum of the generalized eigenvalue problems. Here, the generalized eigenvalue problems is given by\footnote{For our convention, the matrix $\mathbf{A}$ is chosen as $\mathbf{L}_1/i$, and the matrix $\mathbf{B}$ is chosen as $\mathbf{L}_2$, where $\mathbf{L}_1$ and $\mathbf{L}_2$ are the matrix version of the operator $L_1$ and $L_2$, respectively.}
\begin{eqnarray}
    (\mathbf{A}-\omega\mathbf{B})\mathbf{v}=0\, .
\end{eqnarray}
The parameter-dependent matrix $\mathbf{A}-\omega\mathbf{B}$ is known as a matrix pencil. The matrix $\mathbf{B}$ is called the ``mass matrix'' and the matrix $\mathbf{A}$ is called the ``stiffness matrix''. Not like the standard eigenvalue problem, there are many definitions of pseudospectra denoted as $\sigma_\epsilon(\mathbf{A},\mathbf{B})$ for a matrix pencil. We select two primary definitions as follows.

\textit{Definition} 1. If $\mathbf{B}$ is nonsingular, one can define define $\sigma_\epsilon(\mathbf{A},\mathbf{B})=\sigma_\epsilon(\mathbf{B}^{-1}\mathbf{A})$. The boundary of $\sigma_\epsilon(\mathbf{A},\mathbf{B})$ is the $\epsilon^{-1}$ level curve of $\lVert(z-\mathbf{B}^{-1}\mathbf{A})^{-1}\rVert$. Therefore, $\sigma_\epsilon(\mathbf{A},\mathbf{B})$ is defined as 
\begin{eqnarray}
    \sigma_\epsilon(\mathbf{A},\mathbf{B})=\{z\in\mathbb{C}:\lVert(z-\mathbf{B}^{-1}\mathbf{A})^{-1}\rVert>1/\epsilon\}\, .
\end{eqnarray}

\textit{Definition} 2. The boundary of $\sigma_\epsilon(\mathbf{A},\mathbf{B})$ is the $\epsilon^{-1}$ level curve of $\lVert(z\mathbf{B}-\mathbf{A})^{-1}\rVert$. Therefore, $\sigma_\epsilon(\mathbf{A},\mathbf{B})$ is defined as 
\begin{eqnarray}
    \sigma_\epsilon(\mathbf{A},\mathbf{B})=\{z\in\mathbb{C}:\lVert(z\mathbf{B}-\mathbf{A})^{-1}\rVert>1/\epsilon\}\, .
\end{eqnarray}

Note that when the matrix $\mathbf{B}=\mathbf{I}$, these two definitions are equivalent. Similar to the ordinary eigenvalue problem, we still need to convert the general norm (Here, it is the energy norm $ \lVert \cdot \rVert_{E}$ in our case.) to a $2$-norm for calculating the pseudospectrum. If we arrive at the Gram matrix $\mathbf{G}$, which is Hermitian and positive definite, then the norm is given by $\lVert\mathbf{u}\rVert=\lVert\mathbf{W}\mathbf{u}\rVert_2$, where the nonsingular matrix $\mathbf{W}$ is obtained from the Cholesky decomposition of $\mathbf{G}$, i.e., $\mathbf{G}=\mathbf{W}^{\star}\mathbf{W}$. For the first definition of the pseudospectrum, we can compute 
\begin{eqnarray}\label{pseudospectrum_formula}
    \lVert(z-\mathbf{B}^{-1}\mathbf{A})^{-1}\rVert&=&\sup_{\mathbf{u}\neq\mathbf{0}}\frac{\lVert\mathbf{W}(z-\mathbf{B}^{-1}\mathbf{A})^{-1}\mathbf{u}\rVert_2}{\lVert\mathbf{W}\mathbf{u}\rVert_2}=\sup_{\mathbf{u}\neq\mathbf{0}}\frac{\lVert\mathbf{W}(z-\mathbf{B}^{-1}\mathbf{A})^{-1}\mathbf{W}^{-1}\mathbf{u}\rVert_2}{\lVert\mathbf{W}\mathbf{W}^{-1}\mathbf{u}\rVert_2}\nonumber\\
    &=&\lVert\mathbf{W}(z-\mathbf{B}^{-1}\mathbf{A})^{-1}\mathbf{W}^{-1}\rVert_2=\lVert(z-\mathbf{W}\mathbf{B}^{-1}\mathbf{A}\mathbf{W}^{-1})^{-1}\rVert_2\, .
\end{eqnarray}
For the second definition of the pseudospectrum, we compute in the similar way 
\begin{eqnarray}
    \lVert(z\mathbf{B}-\mathbf{A})^{-1}\rVert
    =\lVert\mathbf{W}(z\mathbf{B}-\mathbf{A})^{-1}\mathbf{W}^{-1}\rVert_2=\lVert(z\mathbf{W}\mathbf{B}\mathbf{W}^{-1}-\mathbf{W}\mathbf{A}\mathbf{W}^{-1})^{-1}\rVert_2\, .
\end{eqnarray}
Although the pseudospectrum have different definitions for generalized eigenvalue problems, they share the same spirit. Hence, in this paper, we chose definition $1$ to calculate the pseudospectrum.

\section{The deduction of $\partial_{v}E[\varphi]$} \label{derivation of partail E}
In this appendix, we focus on the deduction of $\partial_{v}E[\varphi]$. It directly can be written as
\begin{eqnarray}\label{first deduction of partial E}
    \partial_{v}E[\varphi] & = &\frac{1}{2} \int_{a}^{b} \mathrm{d} r_{{\star}} \Big[  \partial_{vr_{\star}}\varphi\partial_{r_{\star}}\bar{\varphi} + \partial_{vr_{\star}}\bar{\varphi}\partial_{r_{\star}}\varphi + V\partial_{v}(\varphi\bar{\varphi})  \Big]\, ,
\end{eqnarray}
where $\partial_{vr_{\star}}$ represents $\partial_{v}\partial_{r_{\star}}$. On the other hand, the equations of motion for $\varphi$ and $\bar{\varphi}$ can be written explicitly within $(v,r_{\star})$,
\begin{eqnarray}\label{explicit formualtion of EOM of phi}
    (\partial^{2}_{r_{\star}}+2\partial_{vr_{\star}}-V) \varphi=0\, ,\quad \text{and}\quad (\partial^{2}_{r_{\star}}+2\partial_{vr_{\star}}-V) \bar{\varphi}=0\, .
\end{eqnarray}
By substituting $\partial_{vr_{\star}}\varphi=(V-\partial_{r_{\star}}^2)\varphi/2$ and $\partial_{vr_{\star}}\varphi=(V-\partial_{r_{\star}}^2)\bar{\varphi}/2$ into Eq.(\ref{first deduction of partial E}), we are going to get Eq.(\ref{partial energy v}) as follows
\begin{eqnarray}
     \partial_{v}E[\varphi] & = & \frac{1}{2} \int_{a}^{b} \mathrm{d} r_{{\star}} \Big[  \partial_{vr_{\star}}\varphi\partial_{r_{\star}}\bar{\varphi} + \partial_{vr_{\star}}\bar{\varphi}\partial_{r_{\star}}\varphi + V\partial_{v}(\varphi\bar{\varphi})  \Big]\nonumber\\
     &=&\frac{1}{2} \int_{a}^{b} \mathrm{d} r_{{\star}}\Bigg\{\Big[\frac{1}{2}(V-\partial_{r_{\star}}^2)\varphi\Big]\partial_{r_{\star}}\bar{\varphi}+\Big[\frac{1}{2}(V-\partial_{r_{\star}}^2)\bar{\varphi}\Big]\partial_{r_{\star}}\varphi+ V\partial_{v}(\varphi\bar{\varphi})\Bigg\} \nonumber\\
     &=&\frac{1}{2} \int_{a}^{b} \mathrm{d} r_{\star} \Big[ - \frac{1}{2} \partial_{r_{\star}} ( \partial_{r_{\star}}\varphi\partial_{r_{\star}}\bar{\varphi} ) + \frac{1}{2}V\partial_{r_{\star}}(\varphi\bar{\varphi}) + V\partial_{v}(\varphi\bar{\varphi})   \Big]\nonumber\\
    &=&-\frac{1}{4}|\partial_{r_{\star}}\varphi|^{2} \Big|^{b}_{a}+\frac{1}{4}V|\varphi|^{2} \Big|^{b}_{a} + \int_{a}^{b} \mathrm{d}r_{\star} \Big( \frac{1}{2}V\partial_{v}|\varphi|^{2} - \frac{1}{4} \frac{\mathrm{d}V}{\mathrm{d}r_{\star}}|\varphi|^{2} \Big) \, ,
\end{eqnarray}
where the last line we have use integration by parts, and $|\varphi|^2$ represents $\varphi\bar{\varphi}$. 

\section{The numerical approach of pseudospectrum}\label{numerical_approach_of_pseudospectrum}
In this appendix, we give details associated with the discretization method in order to calculate pseudospectrum. First, we employ the Chebyshev spectral method to discretize the differential operator $L_1$ and $L_2$, resulting in their matrix representations $\mathbf{L}_1$ and $\mathbf{L}_2$. In order to better apply the formulas related to the Chebyshev polynomial which is naturally defined on the interval $[-1,1]$, it is convenient to map the radial coordinate $z\in[0,1]$ into $x\in[-1,1]$ via a linear transformation
\begin{eqnarray}\label{coordinate_transformation_x_sigma}
    x=2z-1\, ,\quad z=\frac{x+1}{2}\, .
\end{eqnarray}

The Chebyshev-Lobatto grid is used in discretization, given resolution $N$, it contains $N+1$ points $\{x_j\}_{j=0}^{N}$ with
\begin{eqnarray}
    x_j=\cos\Big(\frac{j\pi}{N}\Big)\, ,\quad j=0,1,\cdots,N\, .
\end{eqnarray}
The variable $\psi$ is discretized as $\mathbf{\Psi}$ given by
\begin{eqnarray}
    \mathbf{\Psi}=(\psi(x_0),\psi(x_1),\cdots,\psi(x_{N-1}),\psi(x_N))^T\, .
\end{eqnarray}

Furthermore, in order to calculate the pseudospectrum, we should turn the inner product  (\ref{scalar_product}) into the discretized version, i.e.,
\begin{eqnarray}
    \langle\psi_1,\psi_2\rangle_E=\mathbf{\Psi}_1^{\star}\cdot \tilde{\mathbf{G}}^E\cdot \mathbf{\Psi}_2\, ,
\end{eqnarray}
where the symbol $\star$ means the conjugate transpose of a vector. The matrix $\tilde{\mathbf{G}}^E$ is known as the Gram matrix corresponding to the inner product. Finally, we need to talk about an important technical note, i.e., the integral quadrature of product of two functions. Consider that functions $\psi_1$ and $\psi_2$ are approximated by $N+1$ Chebyshev polynomials, a grid of $2N+2$ points is required to achieve numerical accuracy. On a practical level, it can be achieved by evaluating at a double resolution $\bar{N}=2N+1$, and then interpolating down to the original resolution $N$. Essentially, it means that the matrix $\tilde{\mathbf{G}}^E$ is given by
\begin{eqnarray}\label{Gram_matrix_interpolation}
    \tilde{\mathbf{G}}^E=\mathbf{P}^T\cdot \mathbf{G}^{E}\cdot \mathbf{P}\, .
\end{eqnarray}
The matrix $\mathbf{P}$ with size $(\bar{N}+1)\times(N+1)$ is called the interpolation matrix, which can be found as follows
\begin{eqnarray}
    \mathbf{P}_{\bar{i}i}=\frac{1}{c_iN}\Big[1+\sum_{j=1}^N(2-\delta_{j,N})T_j(x_{\bar{i}})T_j(x_i)\Big]\, ,\quad \bar{i}=0,\cdots,\bar{N}\quad \text{and}\quad i=0,\cdots,N\, ,
\end{eqnarray}
where $T_{j}$ is the Chebyshev polynomial of order $j$, and 
\begin{eqnarray}
    c_{i} = \begin{cases}
        2 \, , & i = 0 \quad \text{or} \quad N \, ,\\
        1 \, , & \text{otherwise} .
    \end{cases} 
\end{eqnarray}

Moreover, the matrix $\mathbf{G}^E$ with size $(\bar{N}+1)\times(\bar{N}+1)$ in Eq.(\ref{Gram_matrix_interpolation}) is expressed as
\begin{equation}
    \mathbf{G}^{E}=\mathbf{D}^{T}_{\bar{N}} \cdot \mathbf{C}_{2}\cdot \mathbf{D}_{\bar{N}}+\frac{1}{2}\Big(\mathbf{C}_{1}\cdot\mathbf{D}_{\bar{N}} + \mathbf{D}^{T}_{\bar{N}}\cdot\mathbf{C}_{1}\Big)+\frac14\mathbf{C}_{V}\, ,
\end{equation}
where the matrices $\mathbf{C}_1$, $\mathbf{C}_2$ and $\mathbf{C}_V$ are all diagonal. Their diagonal elements are
\begin{eqnarray}
    (\mathbf{C}_{l})_{\bar{i}\bar{i}}=\frac{2\mu_l(x_{\bar{i}})}{c_{\bar{i}}\bar{N}}\Big[1-\sum_{k=1}^{\lfloor \frac{\bar{N}}{2} \rfloor}T_{2k}(x_{\bar{i}})\frac{2-\delta_{2k,\bar{N}}}{4k^2-1}\Big]\, ,\quad \bar{i}=0,\cdots,\bar{N} \quad \text{and} \quad l=1\, ,2\, ,V\, ,
\end{eqnarray}
where $\lfloor a \rfloor$ is the floor function, i.e., the largest integer that is less than or equal to $a$. Three weight functions $\mu_l$ are 
\begin{eqnarray}
    \mu_{2}(x)&=&\Big(\frac{1+x}{2}\Big)^{2\lambda+2}f
    \Big(\frac{1+x}{2}\Big)\, ,\nonumber\\
    \mu_{1}(x) &=& \lambda\Big(\frac{1+x}{2}\Big)^{2\lambda+1}f\Big(\frac{1+x}{2}\Big)\, ,\nonumber\\
    \mu_{V}(x)&=&\Big(\frac{1+x}{2}\Big)^{2\lambda-2} \Bigg\{\lambda^{2}\Big(\frac{1+x}{2}\Big)^{2} f\Big(\frac{1+x}{2}\Big) + \frac{V[(1+x)/2)]}{f[(1+x)/2]}\Bigg\}\, ,
\end{eqnarray}
respectively. Combined Appendix.\ref{pseudospectrum_of_the_generalized_eigenvalue_problems}, we have obtained the necessities for the pseudospectrum of a generalised eigenvalue problem so far. More specifically, the matrix $\mathbf{G}$ appearing at Appendix.\ref{pseudospectrum_of_the_generalized_eigenvalue_problems} is nothing but the matrix $\tilde{\mathbf{G}}^E$.

\section{The convergence test of QNM spectra}\label{convergence_test}
\begin{figure}[htbp]
    \centering
    \includegraphics{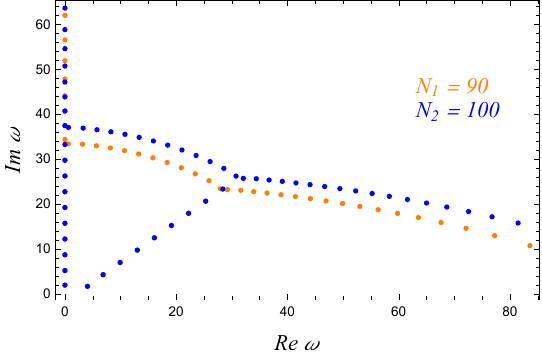}
    \caption{ The discrete eigenvalues calculated with benchmark parameters in the right side of the imaginary axis with different resolutions $N_1 = 90$~(orange) and $N_2 = 100$~(blue).}
    \label{fig:Eigenvalues Compared}
\end{figure}

A simple way to observe the convergence of eigenvalues is to plot the eigenvalues under different resolutions on the same graph. If the eigenvalues are indeed convergent, then these eigenvalues will be highly overlapped under different resolutions. In Fig.\ref{fig:Eigenvalues Compared}, we compare the eigenvalues with the benchmark parameters but from two different resolutions with $N_1=90$ and $N_2=100$. The convergence of QNM spectra for the first few eigenvalues can be demonstrated from Fig.\ref{fig:Eigenvalues Compared}. 

However, in this appendix, we will consider the convergence of eigenvalues from a more sophisticated perspective to avoid misjudgment caused by eyes. According to the Eigenvalue Rule-of-Thumb, numerous eigenvalues of the discretized matrix often bear no meaningful relation to the eigenvalues of the fundamental differential eigen problem, amounting to mere numerical artifacts~\cite{2000Chebyshev}. Motivated by previous statement, we give a convergence test of the QNM spectra in this appendix in the wake of~\cite{2000Chebyshev, BOYD199611}. 

How can we distinguish between the ``good'' and ``bad'' eigenvalues? Considering the feature of the QNM problem, the spectra are symmetric about the imaginary axis. Therefore, we only need to care about the convergence of the eigenvalues on one side of the imaginary axis, including the eigenvalues on the imaginary axis.  Only those eigenvalues that exhibit minimal differences or ``resolution-dependent drift'' can be considered reliable. For two different resolutions $N_1$ and $N_2$ with $N_1<N_2$, we use $\bar{n}_1$ and $\bar{n}_2$ to represent the number of eigenvalues on the right side of the imaginary axis at different resolutions, respectively, and we count the numbers of eigenvalues according to the size of the value of their imaginary parts. For lower resolution $N_1$, it is recommend to define the two kinds of drifts which are defined as
\begin{eqnarray}\label{ordinal_ratios}
    r_{\bar{n},\text{ordinal}}=\frac{\min(|\omega_{\bar{n}}|,\sigma_{\bar{n}})}{\delta_{\bar{n},\text{ordinal}}}\, ,\quad \bar{n}=1,2,\cdots,\bar{n}_1-1,\bar{n}_1\, ,
\end{eqnarray}
\begin{eqnarray}\label{nearest_ratios}
    r_{\bar{n},\text{nearest}}=\frac{\min(|\omega_{\bar{n}}|,\sigma_{\bar{n}})}{\delta_{\bar{n},\text{nearest}}}\, ,\quad \bar{n}=1,2,\cdots,\bar{n}_1-1,\bar{n}_1\, ,
\end{eqnarray}
in which the weight function $\sigma_{\bar{n}}$ is defined as
\begin{eqnarray}
    \sigma_{\bar{n}}=\begin{cases}
        |\omega_1-\omega_2| \, , & \bar{n}=1 \, ,\\
        |\omega_{\bar{n}_1-1}-\omega_{\bar{n}_1}| \, , & \bar{n}=\bar{n}_1 \, , \\
        \frac{1}{2}\Big(|\omega_{\bar{n}-1}-\omega_{\bar{n}}|+|\omega_{\bar{n}+1}-\omega_{\bar{n}}|\Big) \, , & \bar{n}=2,\cdots,\bar{n}_1-1 \, .
    \end{cases}
\end{eqnarray}
The difference between the two definitions origins from the denominator, depending on what pairs of eigenvalues in different resolutions should be compared, the first definition is named as ``ordinal drift''
\begin{eqnarray}\label{ordinal_drift}
    \delta_{\bar{n},\text{ordinal}}=|\omega_{\bar{n}}^{(N_1)}-\omega_{\bar{n}}^{(N_2)}|\, ,
\end{eqnarray}
and the second definition is named as ``nearest neighbor drift''
\begin{eqnarray}\label{nearest_drift}
    \delta_{\bar{n},\text{nearest}}=\min_{k=1,\cdots,\bar{n}_2}|\omega_{\bar{n}}^{(N_1)}-\omega_k^{(N_2)}|\, ,
\end{eqnarray}
where $\omega_j^{(N)}$ is the $j$-th eigenvalue extracted from $\bar{n}_1$ or $\bar{n}_2$ eigenvalues. 

\begin{figure}[htbp]
    \centering
    \includegraphics[width = 0.48\textwidth]{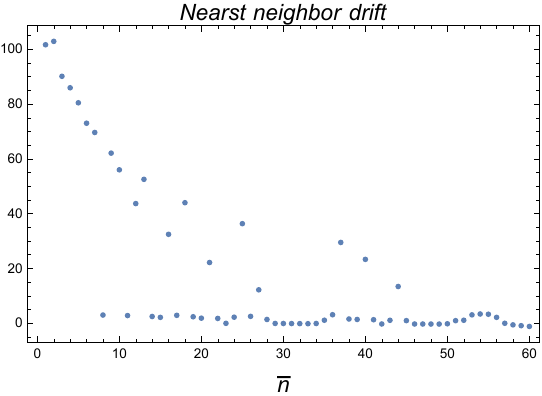}
    \includegraphics[width = 0.48\textwidth]{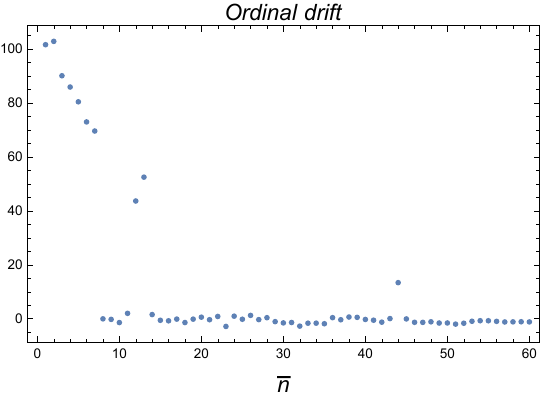}
    \caption{ The eigenvalues are computed with benchmark parameters. The left image represents the result of the nearst neighbor drift, and the right image represents the result of the ordinal drift. We have taken the natural logarithm of $\delta_{\bar{n},\text{ordinal}}$ and $\delta_{\bar{n},\text{nearest}}$ .}
    \label{fig:Spectrum Convergence}
\end{figure}

When the eigenvalues are accurate enough, the two ratios would be very large. In Fig.\ref{fig:Spectrum Convergence}, we show the two ratios (\ref{ordinal_ratios}) and (\ref{nearest_ratios}) with two resolutions $N_{1} = 90$ and $N_{2}= 100$. Note that the scale of the vertical axis has been set to the logarithmic value of the drift. The advantage of the drift ratio measure lies in its resilience against spurious modes stemming from numerical errors, as such errors are unlikely to persist near meaningful eigenvalues as the resolution varies.

\section{The time-symmetric numerical methods}\label{numerical_method}
In this appendix, following the Refs.\cite{markakis2019timesymmetry,OBoyle:2022yhp,Markakis:2023pfh}, we are going to solve the coupled ODEs (\ref{linear_evolution_system}) from the methods of lines for PDE (\ref{equation_of_evolution}). We start from the following linear differential equations
\begin{eqnarray}\label{odes_method}
    \frac{\mathrm{d}\mathbf{\Psi}}{\mathrm{d}v}=\mathbf{H}\mathbf{\Psi}\, ,
\end{eqnarray}
where vector $\mathbf{\Psi}$ is the discrete approximation of the field $\psi$ and $\mathbf{H}$ is time-independent. Using the fundamental theorem of calculus, Eq.(\ref{odes_method}) tells us that $\mathbf{\Psi}(v)$ at time $v_{n}$ and $\mathbf{\Psi}(v)$ at time $v_{n+1}$ have the following relation
\begin{eqnarray}\label{integral_equation}
    \mathbf{\Psi}(v_{n+1})=\mathbf{\Psi}(v_n)+\mathbf{H}\int_{v_n}^{v_{n+1}}\mathbf{\Psi}(v)\mathrm{d}v\, .
\end{eqnarray}
The classical Runge-Kutta method for ODEs is based on $1$-point Taylor expansion. The so-called time-symmetric numerical methods is based on $2$-point Taylor expansion. In fact, the integral in Eq.(\ref{integral_equation}) is computed by using the Lanczos-Dyche formula:
\begin{eqnarray}\label{Lanczos_Dyche_formula}
    \int_{v_n}^{v_{n+1}}\mathbf{\Psi}(v)\mathrm{d}v=\sum_{m=1}^lc_{lm}(\Delta v)^m\Big[\mathbf{\Psi}^{(m-1)}(v_n)+(-1)^{m-1}\mathbf{\Psi}^{(m-1)}(v_{n+1})\Big]+\mathbf{R}_l(\mathbf{\Psi})
\end{eqnarray}
with coefficients given by
\begin{eqnarray}
    c_{lm}=\frac{l!(2l-m)!}{m!(2l)!(l-m)!}
\end{eqnarray}
and the remainder term $\mathbf{R}_l(\mathbf{\Psi})$ given by
\begin{eqnarray}
   \mathbf{R}_l(\mathbf{\Psi})=(-1)^l\frac{(l!)^2}{(2n+1)!(2l)!}(\Delta v)^{2l+1}\mathbf{\Psi}^{(2l)}(v)\, ,\quad v\in[v_n,v_{n+1}]\, .
\end{eqnarray}
For convenience, $\mathbf{\Psi}(v_{n})$ is set to be $\mathbf{\Psi}^n$. Furthermore, by ignoring the remainder $\mathbf{R}_l(\mathbf{\Psi})$ of Eq.(\ref{Lanczos_Dyche_formula}), one will grasp at the generalized Hermite rule. 

For $l=2$ used in this study, Eq.(\ref{Lanczos_Dyche_formula}) yields the Hermite rule, which is accurate to fourth order. Then, a fourth scheme for sloving Eq.(\ref{odes_method}) is found which means that
\begin{eqnarray}\label{Hermite_rule}
    \mathbf{\Psi}^{n+1}=\Big[\mathbf{I}-\frac{\Delta v}{2}\mathbf{H}+\frac{(\Delta v)^2}{12}\mathbf{H}^2\Big]^{-1}\Big[\mathbf{I}+\frac{\Delta v}{2}\mathbf{H}+\frac{(\Delta v)^2}{12}\mathbf{H}^2\Big]\mathbf{\Psi}^n\, .
\end{eqnarray}
Here, the time derivatives $\dot{\mathbf{\Psi}}$ has been replaced by $\mathbf{H}\mathbf{\Psi}$. In order to reduce round-off errors at each time step, the above scheme is transformed into a more suitable form of numerical calculation
\begin{eqnarray}\label{Hermite_rule_reduce_round_off_errors}
    \mathbf{\Psi}^{n+1}=\mathbf{\Psi}^n+\Big[\mathbf{I}-\frac{\Delta v}{2}\mathbf{H}\Big(\mathbf{I}-\frac{\Delta v}{6}\mathbf{H}\Big)\Big]^{-1}(\Delta v\mathbf{H})\mathbf{\Psi}^n \, .
\end{eqnarray}
For the choice of $l\ge3$, the numerical scheme can be found in~\cite{markakis2019timesymmetry,OBoyle:2022yhp,Markakis:2023pfh}.
\bibliography{reference}
\bibliographystyle{apsrev4-1}

\end{document}